\documentclass[sigconf]{acmart} 

\usepackage[a-1b]{pdfx}

\usepackage{graphicx}
\usepackage{balance}
\usepackage{comment}
\usepackage{amsfonts}
\usepackage{color,soul}
\usepackage{hyperref}
\usepackage{subcaption}
\usepackage{xspace}
\usepackage{boxedminipage}
\usepackage{diagbox}
\usepackage{multirow}
\usepackage{paralist}
\usepackage{amsmath}
\usepackage{enumitem}
\usepackage{array}
\usepackage{mdframed}
\usepackage[normalem]{ulem}
\usepackage{tabularx}
\usepackage{pgf,tikz}
\usepackage{pgfplots}
\usepackage{mathrsfs}
\usepackage{diagbox}
\usepackage{footnote}
\usepackage{booktabs}
\usepackage{wrapfig}
\usepackage{url}
\usepackage{amsmath}
\usepackage{newtxmath}
\usepackage{arydshln}
\usepackage[normalem]{ulem}
\useunder{\uline}{\ul}{}

\usepackage{circledsteps}

\makeatletter
\DeclareRobustCommand{\iscircle}{\mathord{\mathpalette\is@circle\relax}}
\newcommand\is@circle[2]{%
  \begingroup
  \sbox\z@{\raisebox{\depth}{$\m@th#1\bigcirc$}}%
  \sbox\tw@{$#1\square$}%
  \resizebox{!}{\ht\tw@}{\usebox{\z@}}%
  \endgroup
}
\makeatother

\usepackage[linesnumbered,ruled,vlined]{algorithm2e}

 \newcolumntype{L}[1]{>{\raggedright\arraybackslash}p{#1}}
\SetCommentSty{algcmmt}
\SetKwInput{KwInput}{Input}                
\SetKwInput{KwOutput}{Output}

\newcommand{\algname}{PDSum}

\AtBeginDocument{%
  \providecommand\BibTeX{{%
    \normalfont B\kern-0.5em{\scshape i\kern-0.25em b}\kern-0.8em\TeX}}}

\copyrightyear{2023}
\acmYear{2023}
\setcopyright{acmlicensed}\acmConference[WWW '23]{Proceedings of the ACM Web Conference 2023}{May 1--5, 2023}{Austin, TX, USA}
\acmBooktitle{Proceedings of the ACM Web Conference 2023 (WWW '23), May 1--5, 2023, Austin, TX, USA}
\acmPrice{15.00}
\acmDOI{10.1145/3543507.3583371}
\acmISBN{978-1-4503-9416-1/23/04}

\begin{document}

\title{PDSum: Prototype-driven Continuous Summarization of Evolving Multi-document Sets Stream}


\author{Susik Yoon}
\affiliation{%
  \institution{UIUC}
}
\email{susik@illinois.edu}

\author{Hou Pong Chan}
\affiliation{%
  \institution{University of Macau}
}
\email{hpchan@um.edu.mo}

\author{Jiawei Han}
\affiliation{%
  \institution{UIUC}
}
\email{hanj@illinois.edu}


\begin{abstract}
Summarizing text-rich documents has been long studied in the literature, but most of the existing efforts have been made to summarize a static and predefined multi-document set. With the rapid development of online platforms for generating and distributing text-rich documents, there arises an urgent need for continuously summarizing dynamically evolving multi-document sets where the composition of documents and sets is changing over time. This is especially challenging as the summarization should be not only effective in incorporating relevant, novel, and distinctive information from each concurrent multi-document set, but also efficient in serving online applications. In this work, we propose a new summarization problem, Evolving Multi-Document sets stream Summarization (EMDS), and introduce a novel unsupervised algorithm \algname{} with the idea of prototype-driven continuous summarization. \algname{} builds a lightweight prototype of each multi-document set and exploits it to adapt to new documents while preserving accumulated knowledge from previous documents. To update new summaries, the most representative sentences for each multi-document set are extracted by measuring their similarities to the prototypes. A thorough evaluation with real multi-document sets streams demonstrates that \algname{} outperforms state-of-the-art unsupervised multi-document summarization algorithms in EMDS in terms of relevance, novelty, and distinctiveness and is also robust to various evaluation settings.
\end{abstract}

\ccsdesc[500]{Information systems~Summarization}
\ccsdesc[500]{Information systems~Web searching and information discovery}
\ccsdesc[500]{Information systems~Data stream mining}

\keywords{Continuous summarization, Evolving multi-document sets, Unsupervised text summarization}

\maketitle

\section{Introduction}
The rapid development of web-based platforms and digital journalism leads to abundant text-rich content generated in real-time such as news articles, blog posts, online reviews, and scientific articles\,\cite{text_survey2, W2E, WCEP}. As the scale and speed of the generated document streams are overwhelming, it is not feasible for people to digest all the documents of interest by themselves. There have been long-standing efforts devoted to transforming such text-rich documents into a brief and concise summary. Existing studies assume documents of interest (e.g., articles of a news story) are given and return summaries with an abstractive or extractive approach\,\cite{MDS-DL, text_survey, text_survey2}. The recent large-scale language models have further improved the quality of automatic summarization \,\cite{MDS-DL, primera, pegasus}.

The existing studies, however, are not sufficient to meet emerging practical needs for summarization, as they typically target a \emph{fixed document set of a single interest}. In practice, it is common for people to have multiple interests together and stay up-to-date on each interest concurrently\,\cite{cep, mdual, cugola2012processing}. Likewise, a user may track \emph{multiple document sets of different interests} and expect their up-to-date summaries, while the user's interests and the corresponding documents change over time. We refer to the summarization task in this dynamic scenario as \emph{Evolving Multi-Document Sets Stream Summarization (EMDS)}.

\begin{figure}[!t]
    \centering
\includegraphics[width=\columnwidth]{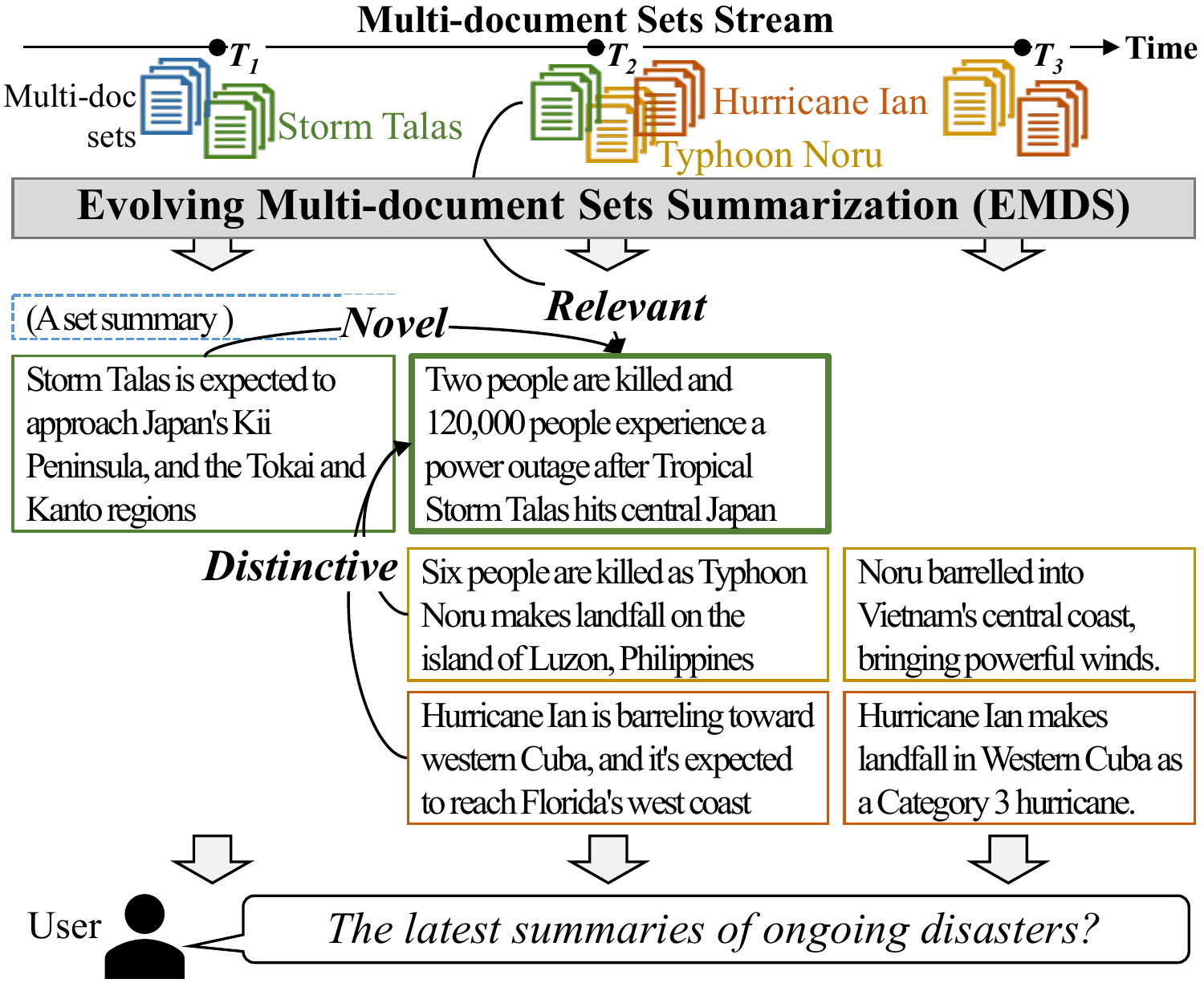}
    \vspace{-0.7cm}
    \caption{An example of continuous summarization of evolving multi-document sets stream (best viewed in color).}
    \label{fig:problem}
    \vspace{-0.3cm}
\end{figure}

Figure \ref{fig:problem} illustrates an example scenario of EMDS. A user would like to get the latest summaries of news stories about disasters around the world at a certain time interval (e.g., every day). The queries specifying particular news stories can be manually issued by the user or automatically provided by an application. For instance, at $T_2$, news articles about \texttt{Storm Talas}, \texttt{Typhoon Noru}, and \texttt{Hurricane Ian} are published. The related articles arrive at varying rates as news stories emerge, continue, and expire. By continuously providing the latest summaries for each news story, the user can easily stay up-to-date with recent news stories of interest. These summaries can be utilized in various downstream tasks such as news curation, event detection, and topic mining\,\cite{newshead, arcus, lee2022taxocom, topicexpan}. 

Despite the practicality and benefits of EMDS, there are unique challenges in EMDS that limit the adoption of existing studies.

\noindent (1) The summarization should consider \emph{documents-, sets- and time-aware themes comprehensively}. For instance, in Figure \ref{fig:problem}, the summary for the story (i.e., set) \texttt{Storm Talas} at $T_2$ should be \emph{relevant} to the current articles (i.e., documents) in the story at $T_2$, \emph{novel} to the previous articles in the story at $T_1$, and \emph{distinctive} from the articles in the other stories. Existing methods typically focus on the relevance of a summary to a target document set but do not consider its novelty compared with the previous documents and/or its distinctiveness to the other sets of documents.

\noindent (2) The summarization should be conducted \emph{with single-pass processing without access to previous documents}. As it is not feasible for online applications to store continuous and unbounded data streams, streaming algorithms typically adopt single-pass processing of data streams for efficiency\,\cite{nets, cugola2012processing, stare}. Similarly in EMDS, once news summaries are derived from the latest document sets, it is more practical to discard them immediately. This makes it more difficult for existing methods to keep track of relevant, novel, and distinctive themes of evolving multi-document sets.

To fill this gap, we propose a novel \emph{unsupervised} summarization method, \textbf{\algname{}} (\textul{P}rototype-\textul{D}riven continuous \textul{Sum}marization for evolving multi-document sets stream), targetting the newly introduced EMDS task. 
\algname{} builds \emph{lightweight prototypes} for multi-document sets to embed and summarize new documents over evolving multi-document sets stream. The set prototypes incorporate the unique \emph{symbolic and semantic themes} of each set and are continuously updated to be distinctive from one another through a contrastive learning objective. Then, in every temporal context, new summaries for each set are derived by extracting the representative sentences in the set prioritized by their symbolic and semantic similarities to the corresponding set prototype. In the meantime, \emph{accumulated knowledge distillation} regularizes the set prototypes to control the balance between the consistent relevance and the novelty of summaries over time.

In summary, the main contributions of this work are as follows:
\vspace{-0.18cm}
\begin{itemize}[leftmargin=12pt, noitemsep]
\item We introduce \textbf{a new summarization problem EMDS} (evolving multi-document sets stream summarization), which is more suitable for dynamic online scenarios than existing summarization tasks and thus is expected to bring huge benefits to users and relevant online applications.
\item We propose \textbf{a novel method \algname{}} designed for EMDS, exploiting lightweight set prototypes learned through contrastive learning with accumulated knowledge distillation. The source code is available at \url{https://github.com/cliveyn/PDSum}.
\item In experiments with real benchmark datasets, \algname{} shows \textbf{the state-of-the-art performance} compared with existing methods in the comprehensive evaluation with \emph{relevance}, \emph{novelty}, and \emph{distinctiveness} measures specifically designed for EMDS.
\end{itemize}

\renewcommand{\arraystretch}{0.2}
\begin{table}[!t]
\small
\caption{A comparison of the proposed summarization task (EMDS) with existing summarization tasks}
\vspace{-0.4cm}
\label{tbl:comparison}
\centering
\begin{tabular}{m{0.5cm}cccc}
\toprule
\multicolumn{1}{l}{} & Multi-doc & Multi-set & Streaming docs & Evolving sets \\  \toprule
SDS & $\times$ & $\times$ & $\times$ & $\times$ \\ \midrule
MDS & $\iscircle$ & $\times$ & $\times$ & $\times$ \\ \midrule
QFS & $\iscircle$ & $\times$ & $\times$ & $\times$ \\ \midrule
TLS & $\iscircle$ & $\times$ & $\times$ & $\times$ \\ \midrule
RTS & $\iscircle$ & $\times$  & $\iscircle$ & $\times$ \\ \midrule
\textbf{EMDS} & \textbf{$\iscircle$} & \textbf{$\iscircle$} & \textbf{$\iscircle$} & \textbf{$\iscircle$}\\ \bottomrule
\end{tabular}
\vspace{-0.5cm}
\end{table}

\section{Related Work} 
\label{sec:related_work}
Text summarization has been actively studied in recent decades\,\cite{text_survey,text_survey2, MDS-DL}. We briefly introduce existing relevant summarization tasks, compared with EMDS in Table \ref{tbl:comparison}, and the representative approaches. 

\vspace{-0.1cm}
\subsection{Sigle/multi-document summarization}
Single document summarization (SDS) and multi-document summarization (MDS) are the most popular summarization tasks, where the former assumes a single document as input while the latter assumes a set of documents of a certain interest (e.g., topic, query, or theme). Typical SDS and MDS methods can be classified as abstractive or extractive approaches, depending on how the summaries are derived, or as supervised or unsupervised approaches, depending on the use of reference (gold) summaries for model training.

Various approaches for SDS and MDS have been studied in the literature. Centroid-based methods\,\cite{MEAD, centroid_revisit, centroid_word} are one of the widely used approaches that cluster input documents and pick the most central sentences as a summary. Graph-based methods embed documents in a graph structure\,\cite{textrank, lexrank, summpip, sumdocs}. A popular method LexRank\,\cite{lexrank} constructs a graph by connecting sentences based on their similarities and applies PageRank\,\cite{pagerank} to extract the most salience sentences. An unsupervised method SummPip\,\cite{summpip} with graph clustering and compression techniques shows comparable performances with supervised methods. Recently, deep neural network (DNN)-based methods have been actively proposed\,\cite{MDS-DL}, where deep reinforcement learning\,\cite{RL-MMR}, semantic text matching\,\cite{MATCHSUM}, hierarchical transformer~\cite{DBLP:conf/acl/HierarchicalTransformer}, or graph neural network\,\cite{HETERSUMGRAPH,DBLP:conf/acl/BASS21,DBLP:conf/acl/Graphsum20} are used, to name a few. While most existing DNN-based methods adopt supervised training with reference summaries, Zhang et al.\,\cite{pegasus} proposed a self-supervised approach with the Gap Sentence Generation objective. PRIMERA\,\cite{primera} further improves the self-supervision by using the Entity Pyramid for masking sentences and provides state-of-the-art pretrained MDS models.

Nevertheless, the existing work for SDS and MDS inherently considers a \emph{static} and \emph{single} set of documents for summarization, which fall too short for continuous summarization of \emph{streaming} documents from \emph{evolving} sets in EMDS. Moreover, some supervised methods require reference summaries which are not readily available in an online scenario.

Another related line of work centers on query-focused summarization (QFS)~\cite{vig2021exploringQfs,DBLP:conf/acl/QueryFreeQfs21,DBLP:conf/sigir/CTSumWanZ14,DBLP:conf/cikm/Wan09topicsum,controlsumTACL21}. 
QFS aims to summarize a fixed set of documents with respect to a user-specified query (e.g., questions, entities, or keywords). 
On the other hand, our EMDS task focuses on summarizing \emph{multiple} document sets of different interests that are \emph{evolving} over time. The summaries in our task should also take into account its distinctiveness to other sets of documents and its novelty to past documents. 

\subsection{Timeline/real-time summarization}
Other relevant tasks are timeline summarization (TLS) and real-time summarization (RTS). TLS generates a timeline of events from a given set of documents on a particular topic, by typically conducting two individual subtasks: date selection and date summarization\,\cite{ETLS}. Graph-based\,\cite{OptimalTransport, graphTLS}, time-event memory-based\,\cite{MTS}, or affinity propagation-based approaches \,\cite{MTLS} have been introduced for TLS. While most of them derive a single timeline from a single document set, a recent work\,\cite{MTLS} generates multiple timelines with different topics. While they are \emph{retrospective} methods for deriving the complete timeline(s) from given documents, EMDS requires continuous updates of summaries from evolving document sets.

RTS\,\cite{trec} (and similar tasks such as temporal summarization\,\cite{TS}, update summarization\,\cite{IUS}, stream summarization\,\cite{BINet}) aims to summarize new documents from document streams considering their relevance, redundancy, and timeliness\,\cite{MDS-DL}. Some of the existing studies model the task as a sequential decision-making problem and apply deep reinforcement learning\,\cite{DRES, NNRL}. Others represent documents in an information network and apply variants of PageRank\,\cite{BINet}. However, their summarization goals are much simpler, predicting a \emph{predefined} relevance labels (e.g., `not relevant', `relevant', or `highly relevant') for documents or sentences stream\,\cite{DRES, NNRL, BINet, TS} retrieved with a \emph{single} and \emph{static} query\,\cite{DRES, NNRL, IUS}. Thus, they are not directly applicable to EMDS.
\section{Problem Setting}

\begin{figure}[!t]
    \centering
\includegraphics[width=0.85\columnwidth]{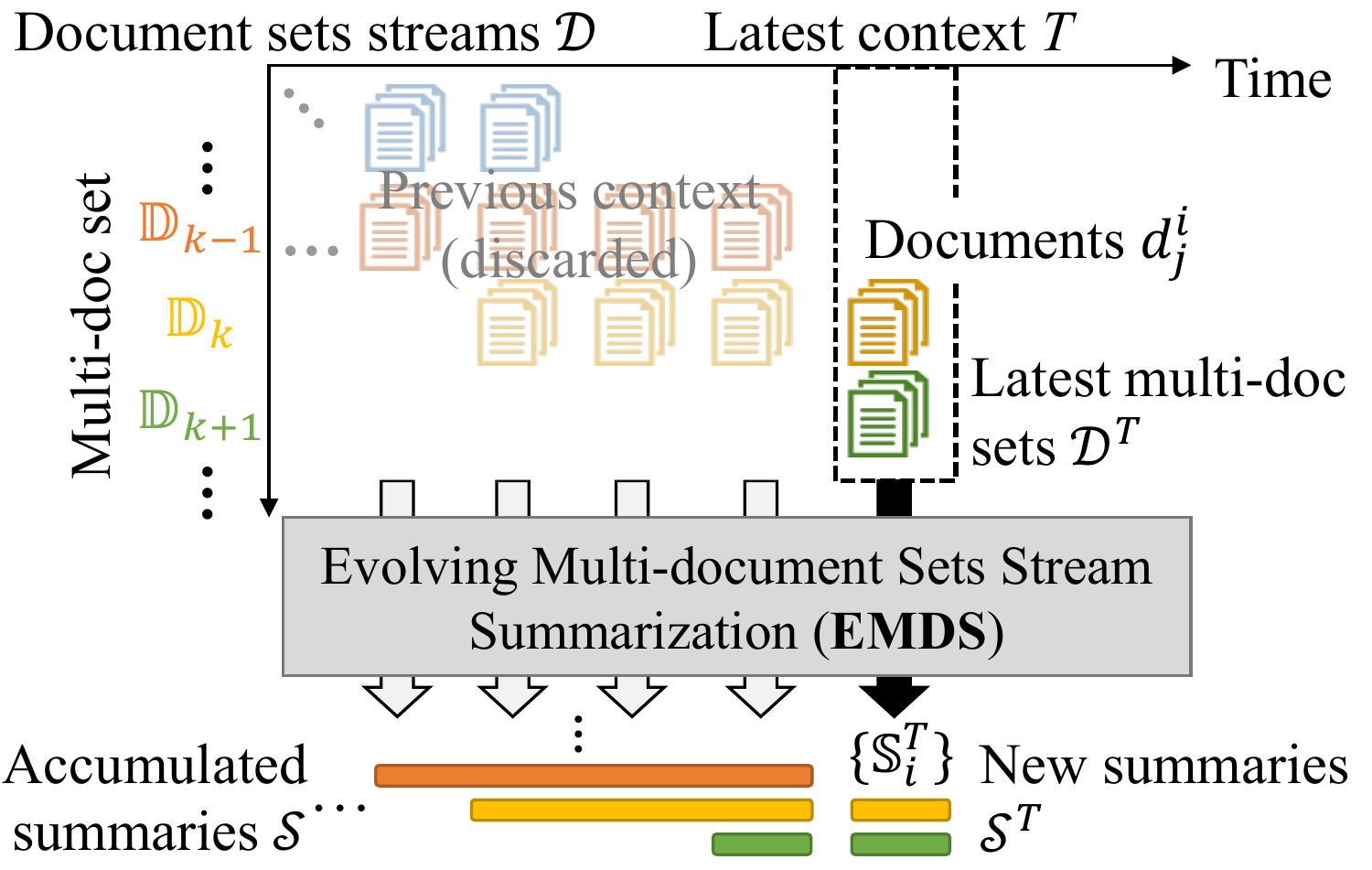}
    \vspace{-0.2cm}
    \caption{An illustration of the EMDS task.}
    \label{fig:definition}
    \vspace{-0.4cm}
\end{figure}

Let a document $d = [s_1, s_2, \ldots, s_{|d|}]$ be a set of sentences $s$ and a multi-document set (or simply \emph{set}) $\mathbb{D} = [d_1, d_2, \ldots, d_{|\mathbb{D}|}]$ be a set of documents under a certain interest (e.g., topic, story, or theme). Then, Definition \ref{def:EMDSS} formally introduces an \emph{evolving multi-document sets stream} considered in this work.

\begin{definition}
\label{def:EMDSS} 
(\textbf{\textsc{Evolving multi-document sets stream}}) An evolving multi-document sets stream $\mathcal{D}=\{\mathbb{D}_i\}$ is composed of unbounded multi-document sets $\mathbb{D}_i$ of documents $d_j^i$. The composition of sets and corresponding documents in $\mathcal{D}$ is continuously changing over time as new documents arrive in new (or existing) sets. A temporal context (or simply \emph{context}) $T$ indicates a certain temporal scope of interest in $\mathcal{D}$. Then, 
$\mathcal{D}^T=\{\mathbb{D}^T_i\}$ represents the sets and corresponding documents arrived at $T$.
\end{definition}

A typical example of evolving multi-document sets stream is a stream of news stories curated by news applications. If a user subscribes to certain categories or topics, news articles (documents) in the relevant news stories (sets) are continuously delivered. Then, the user wants to continuously get new summaries for each ongoing story in every latest context (e.g., every day) of the evolving news stories stream. Definition \ref{def:EMDS} formalizes the summarization task proposed in this work and Figure \ref{fig:definition} illustrates it.

\begin{definition}
\label{def:EMDS} 
(\textbf{\textsc{Evolving Multi-Document sets stream Summarization (EMDS)}}) From evolving multi-document sets stream $\mathcal{D}$, for $\mathcal{D}^T$ in every latest context $T$, a goal of EMDS is to derive new set summaries $\mathcal{S}^T =\{\mathbb{S}^T_i\}$ for each set $\mathbb{D}^T_i \in \mathcal{D}^T$, where $\mathcal{D}^T$ is discarded after being summarized. 
\end{definition}

Please note that EMDS naturally follows an \emph{unsupervised} approach since obtaining reference summaries for multiple sets in different contexts is prohibitively expensive in a streaming setting. In this work, we adopt an \emph{extractive} summarization approach and select representative sentences in each set as a summary. An abstractive approach may cause the hallucination problem\,\cite{hallucination}, which could be more critical when summarizing dynamic and diverse themes of evolving multi-document sets stream. Table \ref{tbl:notation} summarizes the notations frequently used in this paper.

\renewcommand{\arraystretch}{1.0}
\begin{table}[t!]
\small
\centering
\caption{Notations frequently used in this paper.}
\vspace{-0.3cm}
\label{tbl:notation}
\begin{tabular}{cl} \toprule
Notation & Description \\ \midrule
$\mathcal{D}$ & an evolving multi-document sets stream \\
$s, d, \mathbb{D}$ &  a sentence, a document, a multi-document set, \\ 
$\mathbb{S}, \mathcal{S}$ & a set summary, a collection of set summaries \\
$p, \mathbb{P}, \mathcal{P}$ & a phrase, set phrases, a collection of set phrases\\
$R, \mathcal{R}$ & a set prototype, a collection of set prototypes \\
$T$ & a context for summarization \\
$\gamma$ & a distillation ratio \\
\bottomrule
\end{tabular}
\vspace{-0.3cm}
\end{table}
\section{Methodology}
\label{sec:novelty_aware_clustering}
\subsection{Overview}
\subsubsection{\textbf{Main Idea}} A common goal of summarization is to identify a shared theme of documents and derive a concise and informative summary that best describes the theme. Additionally in EMDS, the theme and summary identification should keep up with the evolving contexts of multi-document sets streams; the theme incorporated in a new summary should be not only \emph{relevant} to the target set but also be \emph{novel} compared with previous summaries and \emph{distinctive} from other sets. These goals also need to be achieved efficiently without accessing previous documents.

\begin{figure}[!t]
    \centering
   \includegraphics[width=\columnwidth]{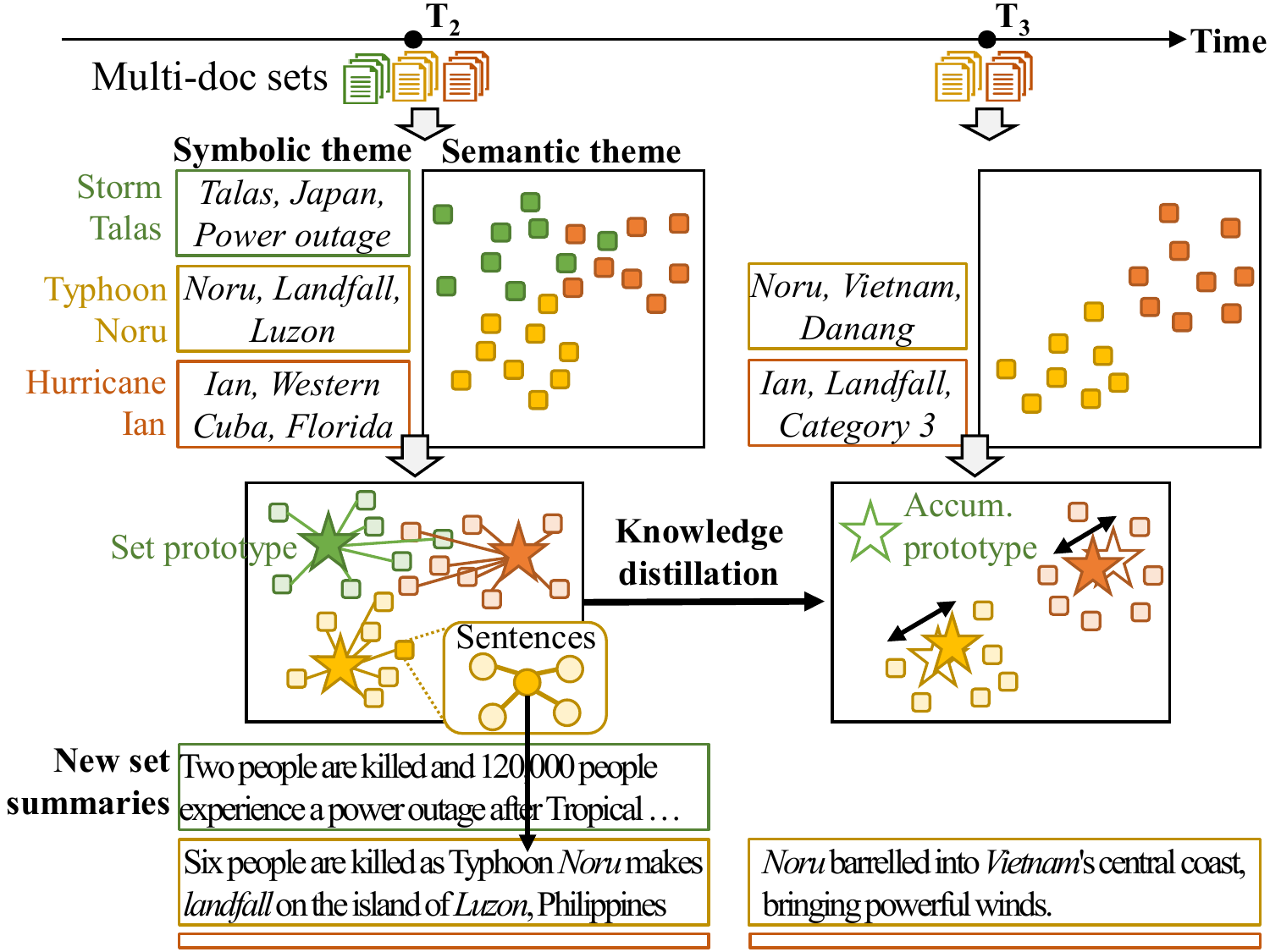}
   \vspace{-0.6cm}
    \caption{Prototype-driven continuous summarization.}
    \label{fig:main_idea}
    \vspace{-0.33cm}
\end{figure}

To this end, we build a lightweight data structure, called \emph{set prototype}, to manage the lifelong theme of a set accumulated over time and use it for efficient and effective continuous summarization. Following the example scenario in Figure \ref{fig:problem}, we illustrate the idea of \textbf{prototype-driven continuous summarization} in Figure \ref{fig:main_idea}.

\noindent (1) First, we identify \emph{the symbolic theme and the semantic theme of a set given concurrent sets}; the former is obtained by identifying top phrases included in documents in each set (i.e., set phrases), while the latter is obtained by representing documents in an embedding space. The two themes complement each other in clarifying unique themes of sets in the current context. For instance, at $T_2$ in Figure \ref{fig:main_idea}, the phrases `\emph{Talas}', `\emph{Japan}', and `\emph{Power outage}' can collectively represent the symbolic theme of \texttt{Storm Talas}. Note that other phrases such as `\emph{Die}', `\emph{Kill}', or `\emph{Approach}' may also be frequently found in the concurrent sets and thus do not represent the unique symbolic themes. These set phrases, however, are sparse and explicit features that can not fully reflect the implicit semantics of documents (e.g., describing victims of disasters). On the other hand, documents of a set represented in an embedding space (marked in rectangles) reflect the semantic themes of the set with dense and contextualized features. The document embeddings of different sets, however, might be overlapped if the documents are written with similar perspectives (e.g., describing evacuation processes for different disasters).

\noindent (2) Thus, we build the prototype of a set by \emph{leveraging the two types of themes together.} Specifically, we consolidate them into set representations (marked in stars in Figure \ref{fig:main_idea}) by averaging document embeddings weighted by set phrases; the documents including more highly-ranked set phrases contribute more to representing the set prototype. Meanwhile, the embedding space is updated to promote documents being closer to their set prototype while being further from the other set prototypes. The embedding learning is regularized through knowledge distillation by balancing the accumulated set prototype (from previous documents) and the new set prototype (from new documents), as shown at $T_3$.

\noindent (3) Then, we identify set summaries as top sentences in each set \emph{prioritized by their similarities to the set prototype.} The prioritization comprehensively considers the document-, sentence-, and phrase-level similarities. As shown in Figure \ref{fig:main_idea}, the chosen summary of \texttt{Typhoon Noru} at $T_2$ is relevant to the set but distinctive from the concurrent sets, and the new summary at $T_3$ conveys novel information while keeping its relevance and distinctiveness.

\begin{figure*}[!t]
    \centering    \includegraphics[width=\textwidth]{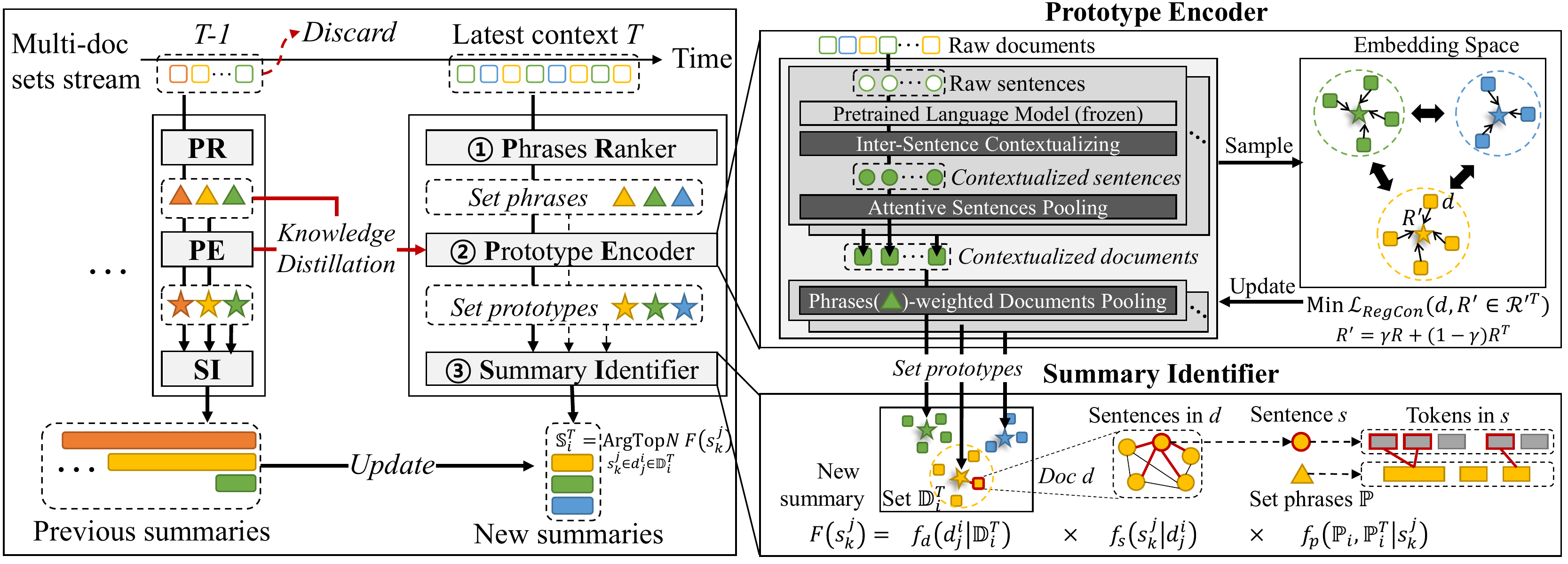}
    \vspace{-0.7cm}
    \caption{The overall procedure of \algname{}. }
    \label{fig:framework}
    \vspace{-0.3cm}
\end{figure*}

\subsubsection{\textbf{Overall Procedure of \algname{}}} We systematically implement the prototype-driven continuous summarization through \algname{}, of which overall procedure is outlined in Algorithm 1 and illustrated in Figure \ref{fig:framework}. In every latest context $T$, the current sets of documents are fed into three sequential components to get new set summaries: (1) \emph{Phrase Ranker} (Line 3) to identify set phrases, (2) \emph{ Prototype Encoder} (Lines 4$-$7 and the upper right part of Figure \ref{fig:framework}) to embed the documents and sets, and (3) \emph{Summary Identifier} (Line 8 and the lower right part of Figure \ref{fig:framework}) to summarize the sets. Each component is explained in detail in the following subsections.


\begin{algorithm}[!t]
    \label{alg:overall}
    \caption{Overall Procedure of \algname{}}
    \small
    \DontPrintSemicolon
    \SetNoFillComment
    \KwInput{an evolving multi-document sets stream $\mathcal{D}$, a distillation ratio $\gamma$, the number $e$ of epochs, the batch size $b$}
    \KwOutput{New summaries $\mathcal{S}^T$ in every context $T$}
    $\mathcal{R},\mathcal{P} \leftarrow \emptyset$ \tcp{Initialize set prototypes and set phrases.}
    \For{every multi-document sets $\mathcal{D}^T$ at the latest context $T$ in $\mathcal{D}$}
    {
        \tcc{Identifying set phrases (Section \ref{sec:phrase_ranker})}
        $\mathcal{P}, \mathcal{P}^T \leftarrow$ PR($\mathbb{D}_i^T \in \mathcal{D}^T$) \\
       \tcc{Encoding documents and sets (Section \ref{sec:prototype_encoder})}
       \For{each epoch in $e$}
        {
           $\mathcal{R},\mathcal{R}^T \leftarrow$ PE$(\mathbb{D}_i^T \in \mathcal{D}^T)$ \\
           $\mathcal{R}^{\prime} \leftarrow \{\gamma R_i + (1-\gamma) R_i^T|R_i \in \mathcal{R}, R_i^T \in \mathcal{R}^T\}$ \\ 
           Update PE with $\sum_{n}\mathcal{L}_{\text{RegCon}}(d\!\in\!\mathcal{D}^T\!,\!\mathcal{R}^{\prime})$ for $|\mathcal{D}^T|/b$ itrs\\
       }
        \tcc{Summarizing sets (Section \ref{sec:summary_identifier})}
        $\mathcal{S}^T \leftarrow$ SI$(\mathbb{D}_i^T \in \mathcal{D}^T,\gamma)$ \\
        Report $\mathcal{S}^T$; \\
    }
\end{algorithm}
\newlength{\oldtextfloatsep}\setlength{\oldtextfloatsep}{\textfloatsep}
\setlength{\textfloatsep}{10pt}

\subsection{Phrase Ranker (PR)}
\label{sec:phrase_ranker}
For each input set, a phrase ranker finds the set phrases representing its symbolic theme in the current context. The set phrases must be found more frequently and uniquely in the set than in any other concurrent set. While any existing phrase mining techniques\,\cite{tfidf, bm25l, UCPhrase} can be adopted, TFIDF\,\cite{tfidf} (default in \algname{}) is a simple but effective choice as it considers term frequencies as well as inverse document frequencies (i.e., inverse set frequencies). By ranking the salience of phrases in a set, the top-$N$ phrases are selected to form set phrases as follows.

\begin{definition}
\label{def:set_phrases} 
(\textbf{\textsc{Set Phrases}}) Given sets $\mathcal{D}^T$ in $T$, set phrases $\mathbb{P}_i^T$ of a set $\mathbb{D}_i^T \in \mathcal{D}^T$ is obtained by a phrase ranker PR($\cdot$):
    \begin{equation}
    \small
    \label{eq:set_phrase}
        \begin{split}
\text{PR}(\mathbb{D}_i^T) = \mathbb{P}_i^T = \{(p_1^i,r_1^i), (p_2^i,r_2^i), \ldots, (p_N^i, r_N^i)\}, \text{where}
        \end{split}
    \end{equation}
phrases $p_k^i$ in $\mathbb{D}_i^T$ are ordered by their score $r_k^i$ (e.g., TFIDF scores).
\end{definition}

\noindent Note that a collection $\mathcal{P}^T=\{\mathbb{P}_i^T\}$ of set phrases in $T$ is continuously added to the accumulated set phrases $\mathcal{P}$ for further use.

\subsection{Prototype Encoder (PE)}
\label{sec:prototype_encoder}
The second component, prototype encoder, conducts two sub-steps. First, it encodes the semantic theme of documents with a pretrained language model and combines them with the identified set phrases to derive set prototypes. Then, it updates the embedding space further to make each set prototype more distinctive from one another while being regularized by accumulated set prototypes.

\subsubsection{\textbf{Encoding Documents and Sets}}
\algname{} first obtains initial sentence representations in each document by using a pretrained sentence encoder (e.g., sentence-BERT\,\cite{sentencebert}). This sentence-level initialization with a pretrained model is more effective and efficient than learning from scratch since the pretrained model has a more generalized embedding capability learned from a much larger and diverse corpus. Furthermore, most sentences meet the maximum input length of typical language models (e.g., 512 tokens).

However, the initialized sentence representations do not reflect the mutual relationships between sentences inside a document since they are embedded independently. Thus, \algname{} further enhances their \emph{inter-sentence contexts} by fine-tuning them with a multi-head self-attention mechanism\,\cite{transformer} as follows.

\begin{definition}
\label{def:con_sent} 
(\textbf{\textsc{Contextualized sentence}}) Given a document $d_j$, the contextualized sentence representations in $d_j$ are:
    \begin{equation}
    \small
    \label{eq:con_sent}
        \begin{split}
            CS(d_j) &= [cs_1^j,\ldots,cs_{|d_j|}^j]\!\in\!\mathbb{R}^{|d_j| \times h_{cs}} \\
            &= l_{\text{ln}}(l_{\text{mhs}}([E(s_k)|s_k\!\in\!d_j])+[E(s_k)|s_k\!\in\!d_j])), \\
        \end{split}
    \end{equation}
where $[E(s_k)|s_k\!\in\!d_j]$ is initial sentence representations by a pretrained language model E($\cdot$), $l_{\text{mhs}}(\cdot)$ is a multi-head self-attention layer, and $l_{\text{ln}}(\cdot)$ is a feed forward layer with layer normalization .
\end{definition}

Then, the contextualized sentence representations are pooled into a contextualized document representation through an \emph{attentive sentences pooling}. This is to verify the relative contribution of each sentence in representing a document. Since the semantics of individual sentences are more diverse than the shared semantics of the document (and that of the set), this helps to filter out noisy sentences (e.g., too specific or too general descriptions) and naturally makes representative sentences stand out to represent the document. Definition \ref{def:con_docs} formalizes the pooling procedure. 

\begin{definition}
\label{def:con_docs} 
(\textbf{\textsc{Contextualized document}}) Given a document $d_j$, a contextualized document representation $cd_j$ is obtained through an attentive pooling of contextualized sentences:
    \begin{equation}
    \small
    \label{eq:con_doc}
        \begin{split}
        CD(d_j) = cd_j &= \sum_{k=1 \ldots |d_j|} \alpha_k cs_k^j \in \mathbb{R}^{h_{cd}} \\
        &= \sum_{k=1 \ldots |d_j|} \frac{\text{exp}([l_{\alpha}(CS(d_j))]_k)}{\sum_{n=1 \ldots |d_j|}\text{exp}([l_{\alpha}(CS(d_j))]_n)}cs_k^j,
        \end{split}
    \end{equation}
where an attention weight $\alpha_k$ indicates the relative importance of $s_k$ for representing $cd_i$, derived by an attention layer $l_{\alpha}(CS(d_j)) = \text{tanh}(CS(d_j)W+b_W)V$ with learnable weights $W, b_W, V$.
\end{definition}

Finally, \algname{} derives a set prototype that represents all documents in the set both in the symbolic and semantic themes, through a \emph{phrase-weighted documents pooling} to combine the contextualized document representations and the accumulated set phrases. Definition \ref{def:set_prototype} formalizes the set prototype.

\begin{definition}
\label{def:set_prototype} 
(\textbf{\textsc{Set Prototype}}) Given a set $\mathbb{D}_i^T$ in $T$, accumulated set phrases $\mathbb{P}_i$, and contextualized document representations $\{cd_j^i\}$, a set prototype $R_i$ derived by a prototype encoder $PE(\cdot)$ is:
    \begin{equation}
    \small
    \label{eq:set_prototype}
        \begin{split}
        PE(\mathbb{D}_i^T) = R_i = \sum_{j=1}^{|\mathbb{D}_i^T|} \Big(\frac{\sum_{(p_k^i,r_k^i) \in \mathbb{P}_i}|p_k^i \in d_j^i|r_k^i}{\sum_{(p_k^i,r_k^i) \in \mathbb{P}_i}|p_k^i \in \mathbb{D}_i^T|r_k^i} \cdot cd_j^i\Big) \in \mathbb{R}^{h_{pe}}.\\
        \end{split}
    \end{equation}
\end{definition}

\subsubsection{\textbf{Optimizing Prototype Encoder}} To optimize the prototype encoder, \algname{} performs \emph{accumulated knowledge distillation} to balance the previously accumulated knowledge and the currently identified new knowledge. Specifically, \algname{} employs two types of set prototype for each set $\mathbb{D}_i^T \in \mathcal{D}^T$: an accumulated set prototype $R_i$ and a new set prototype $R_i^{T}$, where $R_i$ is obtained by Definition \ref{def:set_prototype} while $R_i^{T}$ is obtained similarly but with the \emph{new set phrases} $\mathbb{P}_i^T$ in $T$ and the \emph{initial document representations} as the mean of initial sentence representations: 
\begin{equation}
\small
\label{eq:new_set_prototype}
    \begin{split}
    R_i^{T} = \sum_{j=1}^{|\mathbb{D}_i^T|} \Big(\frac{\sum_{(p_k^i,r_k^i) \in \mathbb{P}_i^T}|p_k^i \in d_j^i|r_k^i}{\sum_{(p_k^i,r_k^i) \in \mathbb{P}_i^T}|p_k^i \in \mathbb{D}_i^T|r_k^i} \cdot \frac{\sum_{s_k^j \in d_j}E(s_k^j)}{|d_j^i|}\Big) \in \mathbb{R}^{h_{pe}},\\
    \end{split}
\end{equation}
In other words, the accumulated set prototype $R_i$ reflects the lifelong theme of a set learned through the previous contexts when the set has existed, whereas the new set prototype $R_i^T$ reflects only the new theme identified in the current context $T$.

Then, a knowledge-distilled set prototype is formulated by combining the two types of set prototypes as follows.

\begin{definition}
\label{def:regularized_set_prototype} 
(\textbf{\textsc{Knowledge-distilled Set Prototype}}) Given an accumulated set prototype $R_i$, a new set prototype $R_i^{T}$, and a distillation ratio $\gamma$, a knowledge-distilled set prototype $R_i^{\prime}$ is:
    \begin{equation}
    \label{eq:reg_set_prototype}
        \begin{split}
        R_i^{\prime} = \gamma R_i + (1-\gamma) R_i^{T},\\
        \end{split}
    \end{equation}
where $\gamma$ controls the distillation of the accumulated theme to the new theme; the higher $\gamma$ weighs more on the previously learned theme while the lower $\gamma$ weighs more on the newly identified theme.
\end{definition}

Finally, \algname{} promotes documents in each set to highlight their own shared theme; the knowledge-distilled set prototype becomes a positive target for the contextualized document representations in the set to be closer to, while those of the other concurrent sets become negative targets to be further from. Definition \ref{def:regularized_contrastive_loss} formalizes the regularized contrastive loss designed to achieve this goal.

\begin{definition}
\label{def:regularized_contrastive_loss} 
(\textbf{\textsc{Regularized Contrastive Loss}}) Given sets $\mathcal{D}^T$, knowledge-distilled set prototypes $\mathcal{R}^{\prime T}$ in $T$, and a temperature scaling value $\tau$, a regularized contrastive loss is calculated as:
\begin{equation}
\small
\label{eq:regularized_contrastive_loss}
    \mathcal{L}_\text{RegCon}(d_j^i\!\in\!\mathcal{D}^T,\!\mathcal{R}^{\prime T}) = -\text{log}\frac{\text{exp}(\text{cos}(cd_j^i,R_i^{\prime}))/\tau}{\sum_{R_k^{\prime} \in \mathcal{R}^{\prime T}}\text{exp}({\text{cos}(cd_j^i,R_k^{\prime})/\tau)}}.
\end{equation}
\end{definition}

\noindent Note that the regularized contrastive loss is \emph{uniquely designed for EMDS}, different from typical contrastive losses\,\cite{contrastive_survey1, contrastive_survey2, simcse}. Rather than optimizing pairwise distances between samples, \algname{} makes the best of set prototypes to achieve a similar goal but in a more efficient and effective way; it significantly reduces the similarity computations (i.e., from $O(N_d^2)$ to $O(N_dN_D)$ for $N_d$ articles and $N_D$ sets where $N_d \gg N_D$) and also directly pursues the unique themes identification of sets (i.e., making articles distinctively similar to the set prototype) which is well aligned with the goal of summarization.

With the accordingly optimized prototype encoder, documents in a set are represented more \emph{relevantly} within the set while retaining more \emph{novelty} from earlier documents in the set and more \emph{distinctiveness} from documents in the other sets.

\subsection{Summary Identifier (SI)}
\label{sec:summary_identifier}
Finally, a summary identifier picks the most representative sentences in each set as a new summary. For each sentence of a document in a set, three levels of the score are conjunctively estimated: (1) the semantic similarity between the set and the document (i.e., doc-level score), (2) the contribution of a sentence in representing the document (i.e., sentence-level score), and (3) the symbolic similarity between the set phrases and the sentence (i.e., phrase-level score). In the meanwhile, the distillation ratio $\gamma$ controls the balance between the accumulated knowledge and the new knowledge for scoring. Definition \ref{def:new_summary} formalizes the summary identifier.

\begin{definition}
\label{def:new_summary} 
(\textbf{\textsc{New Summary}}) Given a set $\mathbb{D}_i^T$ in $T$, a summary identifier derives its new summary SI$(\mathbb{D}_i^T) = \mathbb{S}_i^T$ as the top sentences ranked by a sentence score function $F(s)$:
\begin{equation}
\small
\begin{split}
    &F(s_k^j \in d_j^i \in \mathbb{D}_i^T) = f_d(d_j^i| \mathbb{D}_i^T)\!\times\!f_s(s_k^j|d_j^i)\!\times\!f_p(\mathbb{P}_i,\mathbb{P}_i^T|s_k^j), \text{ where} \\
\end{split}
\end{equation}
\scalebox{0.85}{$
   \begin{aligned}
    \text{\quad}f_d(d_j^i|\mathbb{D}_i^T) &= \gamma \text{exp}(\text{cos}(cd_j^i,R_i)) + (1-\gamma) \text{exp}(\text{cos}(cd_j^i,R_i^T)) \\
    \text{\quad\quad}f_s(s_k^j|d_j^i) &= \alpha_k \text{ (i.e., an attention weight in Equation \ref{eq:con_doc})} \\
    \text{\quad}f_p(\mathbb{P}_i,\mathbb{P}_i^T|s_k^j) &= \gamma \frac{\sum_{(p_k^i,r_k^i) \in \mathbb{P}_i}|p_k^i \in s_k^j|r_k^i}{\sum_{(p_k^i,r_k^i) \in \mathbb{P}_i}|p_k^i \in d_j^i|r_k^i} + (1-\gamma) \frac{\sum_{(p_k^i,r_k^i) \in \mathbb{P}_i^T}|p_k^i \in s_k^j|r_k^i}{\sum_{(p_k^i,r_k^i) \in \mathbb{P}_i^T}|p_k^i \in d_j^i|r_k^i}.
   \end{aligned}$}
\end{definition}

The distillation ratio $\gamma$, utilized for both the embedding learning and the set summarizing, allows a user to have more freedom in choosing a specific degree of knowledge preservation over contexts (e.g., some users may prefer to get fresh information in summaries, while others may want to get more consistent information in summaries across the entire contexts.). In Section \ref{sec:sensitivity}, we study the effects of the distillation ratio and demonstrate that the value of 0.5 balances the trade-off well and results in quality summaries more conforming to the reference summaries provided by humans.

\subsection{Time Complexity of \algname{}}
\label{apx:complexity}
Given $N_d$\,(\# of documents), $N_D$\,(\# of sets), $N_P$\,(\# of set phrases), $N_{PE}$\,(parameter size in a prototype encoder), $N_E$\,(epoch size), and $N_B$\,(batch size), (1) the time complexity for a phrase ranker is $O(N_P N_d)$, (2) that for a prototype encoder is $O(N_d N_{PE} + N_P N_D + N_E N_B N_{PE})$ where $O(N_d N_{PE})$ for embedding documents, $O(N_P N_D)$ for encoding set prototypes, and $O(N_E N_B N_D N_{PE})$ for training, and (3) that for a summary identifier is $O(N_d N_D)$. Since $N_d, N_{PE} \gg N_D, N_P, N_E, N_B$, the total time complexity becomes $O(N_d N_{PE})$.
\section{Experiments}

We conducted extensive experiments to evaluate the performance of \algname{}, of which results are summarized as follows.

\begin{itemize}[leftmargin=10pt, noitemsep]
    \item \algname{} achieved \emph{state-of-the-art performances} in terms of relevance, novelty, and distinctiveness in EMDS on two benchmark datasets compared with existing algorithms (Section \ref{sec:overall_performance}) and returned quality summaries (Appendix \ref{apx:case_study}).
    \item \algname{} was \emph{robust} to the distillation ratio, the number of set phrases, and training hyperparameters (Section \ref{sec:sensitivity}).
    \item \algname{} was more \emph{efficient} than existing algorithms and \emph{scalable} in various streaming settings (Section \ref{sec:scalability}).
\end{itemize}

\begin{table*}[t]
\caption{Overall performance results. The highest and the second highest results are bolded and underlined, respectively.}
\vspace{-0.4cm}
\label{tbl:overall_performance}
\centering
\small
\setlength{\tabcolsep}{4.5pt}
\begin{tabular}{ccccccccc|cccccccc}
\cline{2-17}
 & \multicolumn{8}{c|}{WCEP} & \multicolumn{8}{c}{W2E} \\ \cline{2-17} 
  & \multicolumn{4}{c|}{Relevance} & \multicolumn{2}{c|}{Novelty} & \multicolumn{2}{c|}{Distinctiveness} & \multicolumn{4}{c|}{Relevance} & \multicolumn{2}{c|}{Novelty} & \multicolumn{2}{c}{Distinctiveness} \\ \cline{2-17}
 & R1 & R2 & RL & \multicolumn{1}{c|}{BS} & N-RL & \multicolumn{1}{c|}{N-BS} & D-RL & D-BS & R1 & R2 & RL & \multicolumn{1}{c|}{BS} & N-RL & \multicolumn{1}{c|}{N-BS} & D-RL & D-BS \\ \hline
DocCent & 20.06 & 5.20 & 16.02 & \multicolumn{1}{c|}{83.59} & 10.67 & \multicolumn{1}{c|}{79.77} & 1.19 & 1.95 & 16.06 & 3.88 & 13.06 & \multicolumn{1}{c|}{82.96} & 5.86 & \multicolumn{1}{c|}{78.65} & \uline{1.13} & 1.17 \\
IncDocCent & 20.52 & 5.40 & 16.34 & \multicolumn{1}{c|}{83.73} & 10.51 & \multicolumn{1}{c|}{79.11} & 1.19 & 1.97 & 14.65 & 2.97 & 11.39 & \multicolumn{1}{c|}{82.73} & 5.16 & \multicolumn{1}{c|}{77.89} & 1.11 & 1.14 \\
SentCent & 19.17 & 4.87 & 15.37 & \multicolumn{1}{c|}{83.38} & 10.30 & \multicolumn{1}{c|}{79.35} & 1.18 & 1.92 & 14.93 & 3.41 & 12.15 & \multicolumn{1}{c|}{82.78} &  5.91 & \multicolumn{1}{c|}{\textbf{79.24}} & 1.12 & 1.16 \\
IncSentCent & 19.64 & 4.99 & 15.64 & \multicolumn{1}{c|}{83.46} & 10.53 & \multicolumn{1}{c|}{79.24} & 1.19 & 1.93 & 14.94 & 3.39 & 12.19 & \multicolumn{1}{c|}{82.81} & 5.76 & \multicolumn{1}{c|}{ {\ul 79.19}} & 1.12 & 1.15 \\
Lexrank & 13.63 & 3.33 & 11.28 & \multicolumn{1}{c|}{82.01} & 9.11 & \multicolumn{1}{c|}{80.58} & 1.13 & 1.79 & 11.40 & 2.93 & 9.55 & \multicolumn{1}{c|}{81.24} & 5.12 & \multicolumn{1}{c|}{77.27} & 1.09 &  1.12 \\
Summpip & 21.72 &  7.04 & 17.28 & \multicolumn{1}{c|}{83.60} &  11.74 & \multicolumn{1}{c|}{81.21} & 1.21 & 1.95 & 16.12 & 4.83 & 12.76 & \multicolumn{1}{c|}{82.68} & 5.77 & \multicolumn{1}{c|}{78.39} & 1.12 & 1.18 \\
PRIMERA & 20.87 & 6.84 & 18.45 & \multicolumn{1}{c|}{ 83.84} & 11.65 & \multicolumn{1}{c|}{80.70} & 1.23 &  \uline{1.99} &  16.28 & 5.69 &  14.24 & \multicolumn{1}{c|}{ 83.00} & 3.67 & \multicolumn{1}{c|}{75.61} & 1.11 & 1.13 \\ \hline
\textbf{PDSum} & \textbf{\begin{tabular}[c]{@{}c@{}}26.62\\[-0.5em] \begin{tabular}[c]{@{}c@{}}\scriptsize ${\pm 0.05}$\end{tabular}\end{tabular}} & \textbf{\begin{tabular}[c]{@{}c@{}}9.20\\[-0.5em] \begin{tabular}[c]{@{}c@{}}\scriptsize ${\pm 0.04}$\end{tabular}\end{tabular}} & \textbf{\begin{tabular}[c]{@{}c@{}}21.01\\[-0.5em] \begin{tabular}[c]{@{}c@{}}\scriptsize ${\pm 0.05}$\end{tabular}\end{tabular}} & \multicolumn{1}{c|}{\textbf{\begin{tabular}[c]{@{}c@{}}84.51\\[-0.5em] \begin{tabular}[c]{@{}c@{}}\scriptsize ${\pm 0.01}$\end{tabular}\end{tabular}}} & \textbf{\begin{tabular}[c]{@{}c@{}}12.67\\[-0.5em] \begin{tabular}[c]{@{}c@{}}\scriptsize ${\pm 0.03}$\end{tabular}\end{tabular}} & \multicolumn{1}{c|}{\begin{tabular}[c]{@{}c@{}}{\ul 81.27}\\[-0.5em] \begin{tabular}[c]{@{}c@{}}\scriptsize ${\pm 0.04}$\end{tabular}\end{tabular}} & \begin{tabular}[c]{@{}c@{}}\textbf{1.27}\\[-0.5em] \begin{tabular}[c]{@{}c@{}}\scriptsize ${\pm 0.00}$\end{tabular}\end{tabular} & \textbf{\begin{tabular}[c]{@{}c@{}}\textbf{2.08}\\[-0.5em] \begin{tabular}[c]{@{}c@{}}\scriptsize ${\pm 0.00}$\end{tabular}\end{tabular}} & \textbf{\begin{tabular}[c]{@{}c@{}}19.61\\[-0.5em] \begin{tabular}[c]{@{}c@{}}\scriptsize ${\pm 0.10}$\end{tabular}\end{tabular}} & \textbf{\begin{tabular}[c]{@{}c@{}}6.48\\[-0.5em] \begin{tabular}[c]{@{}c@{}}\scriptsize ${\pm 0.06}$\end{tabular}\end{tabular}} & \textbf{\begin{tabular}[c]{@{}c@{}}15.75\\[-0.5em] \begin{tabular}[c]{@{}c@{}}\scriptsize ${\pm 0.09}$\end{tabular}\end{tabular}} & \multicolumn{1}{c|}{\textbf{\begin{tabular}[c]{@{}c@{}}83.26\\[-0.5em] \begin{tabular}[c]{@{}c@{}}\scriptsize ${\pm 0.02}$\end{tabular}\end{tabular}}} & \begin{tabular}[c]{@{}c@{}}\uline{6.60}\\[-0.5em] \begin{tabular}[c]{@{}c@{}}\scriptsize ${\pm 0.03}$\end{tabular}\end{tabular} & \multicolumn{1}{c|}{\begin{tabular}[c]{@{}c@{}}76.35\\[-0.5em] \begin{tabular}[c]{@{}c@{}}\scriptsize ${\pm 0.09}$\end{tabular}\end{tabular}} & \begin{tabular}[c]{@{}c@{}}\textbf{1.17}\\[-0.5em] \begin{tabular}[c]{@{}c@{}}\scriptsize ${\pm 0.00}$\end{tabular}\end{tabular} & \begin{tabular}[c]{@{}c@{}}\textbf{1.28}\\[-0.5em] \begin{tabular}[c]{@{}c@{}}\scriptsize ${\pm 0.00}$\end{tabular}\end{tabular} \\[-0.2em] \hline 

w/o doc-score & \begin{tabular}[c]{@{}c@{}}{\ul 26.44}\\[-0.5em] \begin{tabular}[c]{@{}c@{}}\scriptsize ${\pm 0.07}$\end{tabular}\end{tabular} & \begin{tabular}[c]{@{}c@{}}{\ul 9.13}\\[-0.5em] \begin{tabular}[c]{@{}c@{}}\scriptsize ${\pm 0.04}$\end{tabular}\end{tabular} & \begin{tabular}[c]{@{}c@{}}{\ul 20.85}\\[-0.5em] \begin{tabular}[c]{@{}c@{}}\scriptsize ${\pm 0.06}$\end{tabular}\end{tabular} & \multicolumn{1}{c|}{\begin{tabular}[c]{@{}c@{}}{\ul 84.47}\\[-0.5em] \begin{tabular}[c]{@{}c@{}}\scriptsize ${\pm 0.01}$\end{tabular}\end{tabular}} & \begin{tabular}[c]{@{}c@{}} {\ul 12.64} \\[-0.5em] \begin{tabular}[c]{@{}c@{}}\scriptsize ${\pm 0.05}$\end{tabular}\end{tabular} & \multicolumn{1}{c|}{\begin{tabular}[c]{@{}c@{}}\textbf{81.40}\\[-0.5em] \begin{tabular}[c]{@{}c@{}}\scriptsize ${\pm 0.04}$\end{tabular}\end{tabular}} & \begin{tabular}[c]{@{}c@{}}\uline{1.26}\\[-0.5em] \begin{tabular}[c]{@{}c@{}}\scriptsize ${\pm 0.00}$\end{tabular}\end{tabular} & \begin{tabular}[c]{@{}c@{}}\textbf{2.08}\\[-0.5em] \begin{tabular}[c]{@{}c@{}}\scriptsize ${\pm 0.00}$\end{tabular}\end{tabular} & \begin{tabular}[c]{@{}c@{}}{\ul 19.07}\\[-0.5em] \begin{tabular}[c]{@{}c@{}}\scriptsize ${\pm 0.01}$\end{tabular}\end{tabular} & \begin{tabular}[c]{@{}c@{}}{\ul 6.23}\\[-0.5em] \begin{tabular}[c]{@{}c@{}}\scriptsize ${\pm 0.05}$\end{tabular}\end{tabular} & \begin{tabular}[c]{@{}c@{}}{\ul 15.31}\\[-0.5em] \begin{tabular}[c]{@{}c@{}}\scriptsize ${\pm 0.08}$\end{tabular}\end{tabular} & \multicolumn{1}{c|}{\begin{tabular}[c]{@{}c@{}}{\ul 83.20}\\[-0.5em] \begin{tabular}[c]{@{}c@{}}\scriptsize ${\pm 0.02}$\end{tabular}\end{tabular}} & \begin{tabular}[c]{@{}c@{}}6.56\\[-0.5em] \begin{tabular}[c]{@{}c@{}}\scriptsize ${\pm 0.03}$\end{tabular}\end{tabular} & \multicolumn{1}{c|}{\begin{tabular}[c]{@{}c@{}}76.60\\[-0.5em] \begin{tabular}[c]{@{}c@{}}\scriptsize ${\pm 0.08}$\end{tabular}\end{tabular}} & \begin{tabular}[c]{@{}c@{}}\textbf{1.17}\\[-0.5em] \begin{tabular}[c]{@{}c@{}}\scriptsize ${\pm 0.00}$\end{tabular}\end{tabular} & \begin{tabular}[c]{@{}c@{}}\uline{1.27}\\[-0.5em] \begin{tabular}[c]{@{}c@{}}\scriptsize ${\pm 0.00}$\end{tabular}\end{tabular} \\[-0.2em]

w/o sent-score & \begin{tabular}[c]{@{}c@{}}22.75\\[-0.5em] \begin{tabular}[c]{@{}c@{}}\scriptsize ${\pm 0.01}$\end{tabular}\end{tabular} & \begin{tabular}[c]{@{}c@{}}7.42\\[-0.5em] \begin{tabular}[c]{@{}c@{}}\scriptsize ${\pm 0.01}$\end{tabular}\end{tabular} & \begin{tabular}[c]{@{}c@{}}17.81\\[-0.5em] \begin{tabular}[c]{@{}c@{}}\scriptsize ${\pm 0.01}$\end{tabular}\end{tabular} & \multicolumn{1}{c|}{\begin{tabular}[c]{@{}c@{}}83.60\\[-0.5em] \begin{tabular}[c]{@{}c@{}}\scriptsize ${\pm 0.00}$\end{tabular}\end{tabular}} & \begin{tabular}[c]{@{}c@{}}11.27\\[-0.5em] \begin{tabular}[c]{@{}c@{}}\scriptsize ${\pm 0.01}$\end{tabular}\end{tabular} & \multicolumn{1}{c|}{\begin{tabular}[c]{@{}c@{}}80.47\\[-0.5em] \begin{tabular}[c]{@{}c@{}}\scriptsize ${\pm 0.03}$\end{tabular}\end{tabular}} & \begin{tabular}[c]{@{}c@{}}1.22\\[-0.5em] \begin{tabular}[c]{@{}c@{}}\scriptsize ${\pm 0.00}$\end{tabular}\end{tabular} & \begin{tabular}[c]{@{}c@{}}1.95\\[-0.5em] \begin{tabular}[c]{@{}c@{}}\scriptsize ${\pm 0.00}$\end{tabular}\end{tabular} & \begin{tabular}[c]{@{}c@{}}14.22\\[-0.5em] \begin{tabular}[c]{@{}c@{}}\scriptsize ${\pm 0.02}$\end{tabular}\end{tabular} & \begin{tabular}[c]{@{}c@{}}4.02\\[-0.5em] \begin{tabular}[c]{@{}c@{}}\scriptsize ${\pm 0.00}$\end{tabular}\end{tabular} & \begin{tabular}[c]{@{}c@{}}11.38\\[-0.5em] \begin{tabular}[c]{@{}c@{}}\scriptsize ${\pm 0.01}$\end{tabular}\end{tabular} & \multicolumn{1}{c|}{\begin{tabular}[c]{@{}c@{}}81.26\\[-0.5em] \begin{tabular}[c]{@{}c@{}}\scriptsize ${\pm 0.00}$\end{tabular}\end{tabular}} & \begin{tabular}[c]{@{}c@{}}\textbf{7.37}\\[-0.5em] \begin{tabular}[c]{@{}c@{}}\scriptsize ${\pm 0.00}$\end{tabular}\end{tabular} & \multicolumn{1}{c|}{\begin{tabular}[c]{@{}c@{}}74.27\\[-0.5em] \begin{tabular}[c]{@{}c@{}}\scriptsize ${\pm 0.07}$\end{tabular}\end{tabular}} & \begin{tabular}[c]{@{}c@{}}1.11\\[-0.5em] \begin{tabular}[c]{@{}c@{}}\scriptsize ${\pm 0.00}$\end{tabular}\end{tabular} & \begin{tabular}[c]{@{}c@{}}1.21\\[-0.5em] \begin{tabular}[c]{@{}c@{}}\scriptsize ${\pm 0.00}$\end{tabular}\end{tabular} \\[-0.2em]

w/o phrase-score & \begin{tabular}[c]{@{}c@{}}20.99\\[-0.5em] \begin{tabular}[c]{@{}c@{}}\scriptsize ${\pm 0.09}$\end{tabular}\end{tabular} & \begin{tabular}[c]{@{}c@{}}6.33\\[-0.5em] \begin{tabular}[c]{@{}c@{}}\scriptsize ${\pm 0.05}$\end{tabular}\end{tabular} & \begin{tabular}[c]{@{}c@{}}17.06\\[-0.5em] \begin{tabular}[c]{@{}c@{}}\scriptsize ${\pm 0.06}$\end{tabular}\end{tabular} & \multicolumn{1}{c|}{\begin{tabular}[c]{@{}c@{}}83.59\\[-0.5em] \begin{tabular}[c]{@{}c@{}}\scriptsize ${\pm 0.02}$\end{tabular}\end{tabular}} & \begin{tabular}[c]{@{}c@{}}11.2\\[-0.5em] \begin{tabular}[c]{@{}c@{}}\scriptsize ${\pm 0.05}$\end{tabular}\end{tabular} & \multicolumn{1}{c|}{\begin{tabular}[c]{@{}c@{}}80.71\\[-0.5em] \begin{tabular}[c]{@{}c@{}}\scriptsize ${\pm 0.08}$\end{tabular}\end{tabular}} & \begin{tabular}[c]{@{}c@{}}1.21\\[-0.5em] \begin{tabular}[c]{@{}c@{}}\scriptsize ${\pm 0.00}$\end{tabular}\end{tabular} & \begin{tabular}[c]{@{}c@{}}1.96\\[-0.5em] \begin{tabular}[c]{@{}c@{}}\scriptsize ${\pm 0.00}$\end{tabular}\end{tabular} & \begin{tabular}[c]{@{}c@{}}13.59\\[-0.5em] \begin{tabular}[c]{@{}c@{}}\scriptsize ${\pm 0.19}$\end{tabular}\end{tabular} & \begin{tabular}[c]{@{}c@{}}3.94\\[-0.5em] \begin{tabular}[c]{@{}c@{}}\scriptsize ${\pm 0.11}$\end{tabular}\end{tabular} & \begin{tabular}[c]{@{}c@{}}10.91\\[-0.5em] \begin{tabular}[c]{@{}c@{}}\scriptsize ${\pm 0.17}$\end{tabular}\end{tabular} & \multicolumn{1}{c|}{\begin{tabular}[c]{@{}c@{}}81.86\\[-0.5em] \begin{tabular}[c]{@{}c@{}}\scriptsize ${\pm 0.04}$\end{tabular}\end{tabular}} & \begin{tabular}[c]{@{}c@{}}5.32\\[-0.5em] \begin{tabular}[c]{@{}c@{}}\scriptsize ${\pm 0.04}$\end{tabular}\end{tabular} & \multicolumn{1}{c|}{\begin{tabular}[c]{@{}c@{}}76.16\\[-0.5em] \begin{tabular}[c]{@{}c@{}}\scriptsize ${\pm 0.33}$\end{tabular}\end{tabular}} & \begin{tabular}[c]{@{}c@{}}1.10\\[-0.5em] \begin{tabular}[c]{@{}c@{}}\scriptsize ${\pm 0.00}$\end{tabular}\end{tabular} & \begin{tabular}[c]{@{}c@{}}1.13\\[-0.5em] \begin{tabular}[c]{@{}c@{}}\scriptsize ${\pm 0.00}$\end{tabular}\end{tabular} \\[-0.2em] \hline
\end{tabular}
\end{table*}

 \begin{figure*}[!t]
     \centering
     \vspace{-0.37cm}
     \includegraphics[width=\textwidth]{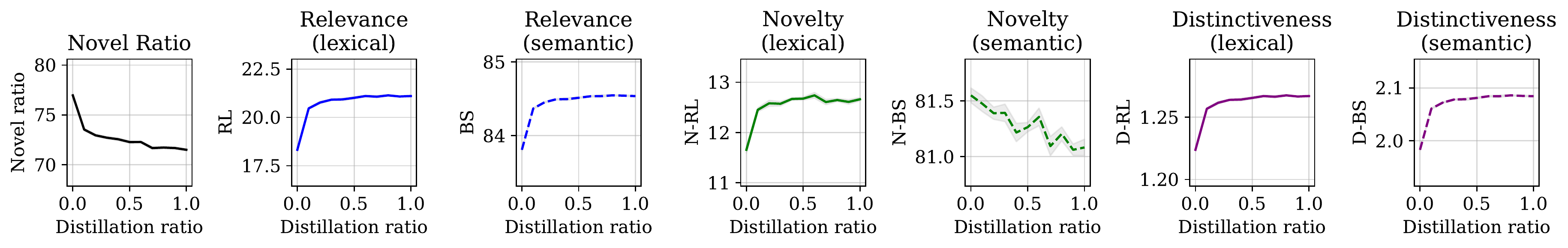}
     \vspace{-0.85cm}
    \caption{Effects of knowledge distillation ratio in WCEP (the result of W2E is provided in Appendix \ref{apx:W2E_RR}).}
    \vspace{-0.4cm}
    \label{fig:WCEP19_RR}
 \end{figure*}


\subsection{Experiment Setting}
\label{exp:setting}
\subsubsection{\textbf{Multi-document Sets Stream}} We used two real news datasets: WCEP\,\cite{WCEP} and W2E\,\cite{W2E}. To the best of our knowledge, they are the only benchmark summarization datasets suitable for EMDS; they contain timestamped news articles (i.e., documents) of different stories (i.e., sets) over various temporal contexts (i.e., evolving sets) and provide reference summaries annotated by humans for evaluation. We simulated the datasets as multi-document sets stream $\mathcal{D}$ and set contexts $T$ according to reference summaries. Refer to Appendix \ref{apx:dataset} for more details.


\subsubsection{\textbf{Compared Algorithms}} For the new task EMDS, we prepared strong baselines by adopting a centroid-based model\,\cite{centroid_word, centroid_revisit} with a pretrained language model: \emph{DocCent} and \emph{SentCent} with document- and sentence-based centers, respectively, and their incremental versions \emph{IncDocCent}, and \emph{IncSentCent}. We also compared three popular \emph{unsupervised} algorithms for multi-document summarization: the graph-based model \emph{Lexrank}\,\cite{lexrank}, the state-of-the-art extractive model \emph{Summpip}\,\cite{summpip}, and the state-of-the-art abstractive model \emph{PRIMERA}\,\cite{primera}. We fed each document set in a context to them so that they can only infer the temporally correlated documents to update their set summaries. See Appendix \ref{apx:implementation} for details.

\subsubsection{\textbf{Evaluation Metrics and Criteria}} We used two popular metrics for evaluating summarization tasks: ROUGE scores\,\cite{rouge} (denoted as \emph{R1}, \emph{R2}, and \emph{RL}) for lexical matching and BERTScore\,\cite{bertscore} (denoted as \emph{BS}) for semantic matching between output summaries and reference summaries. For a more comprehensive evaluation of EMDS, we derive three scores respectively on the two metrics to evaluate output summaries with the following three criteria (refer to Appendix \ref{apx:evaluation} for detailed formulation): 
\begin{itemize}[leftmargin=10pt, noitemsep]
\item \uline{Relevance} between an output summary and a reference summary (i.e., denoted by \emph{RL} and \emph{BS} as default).
\item \uline{Novelty} of an output summary from the previous summary in the same set (i.e., \emph{N-RL} and \emph{N-BS}).
\item \uline{Distinctiveness} between an output summary and the reference summaries of the other concurrent sets (i.e., \emph{D-RL} and \emph{D-BS}). 
\end{itemize}
\vspace{-0.1cm}
For each score, the result averaged over stories and contexts is reported to show the overall performance over the entire stream. 

\subsection{Overall Performance Results}
\label{sec:overall_performance}
Table \ref{tbl:overall_performance} shows the overall evaluation results. For brevity, we show the R1 and R2 scores only for relevance, while the others show similar trends. The main observations are summarized as follows:
\vspace{-0.1cm}
\begin{itemize}[leftmargin=10pt, noitemsep]
    \item \uline{Comparison of existing algorithms:} Among centroid-based variants, IncDocCent achieved the highest scores, indicating the efficacy of document-level and incremental accumulation of knowledge in EMDS. The existing algorithms outperformed them by considering more comprehensive aspects of each set with graph embedding (Summpip) or pretrained Entity Pyramid (PRIMERA). However, since they do not consider previous and concurrent documents conjunctively, their summaries are biased toward the current context and/or the documents in the same set.
    \item \uline{\algname{} v.s. existing algorithms:} \algname{} outperformed existing algorithms in most cases, by achieving significantly higher relevance scores ($\triangle30.9\%$), much higher novelty scores ($\triangle22.0\%$), and moderately higher distinctiveness scores ($\triangle7.2\%$) when averaged over all cases. This indicates the efficacy of set prototypes in identifying distinctive knowledge in concurrent sets and preserving it over continuous sets stream (note that on average 15.49 (std: 7.56) sets existed in each context in the datasets).
    \item \uline{Ablation study for \algname{}:} Among the three levels of scores to prioritize sentences for summarization, the phrase-level score contributed the most as it directly affects the symbolic theme of a set. On the other hand, the document-level score contributed marginally because the prototype encoder makes the representations of documents in the same set converge around the set prototype. Nevertheless, as not all sentences in the document are equally important to represent the set theme, the sentence-level score contributed more to the overall performance.
\end{itemize}
\vspace{-0.1cm}
Appendix \ref{apx:case_study} also discusses human evaluation results and a qualitative case study with sample news stories.

\begin{figure*}[!t]
     \centering
     \begin{subfigure}[b]{0.32\textwidth}
         \centering
         \includegraphics[width=\textwidth]{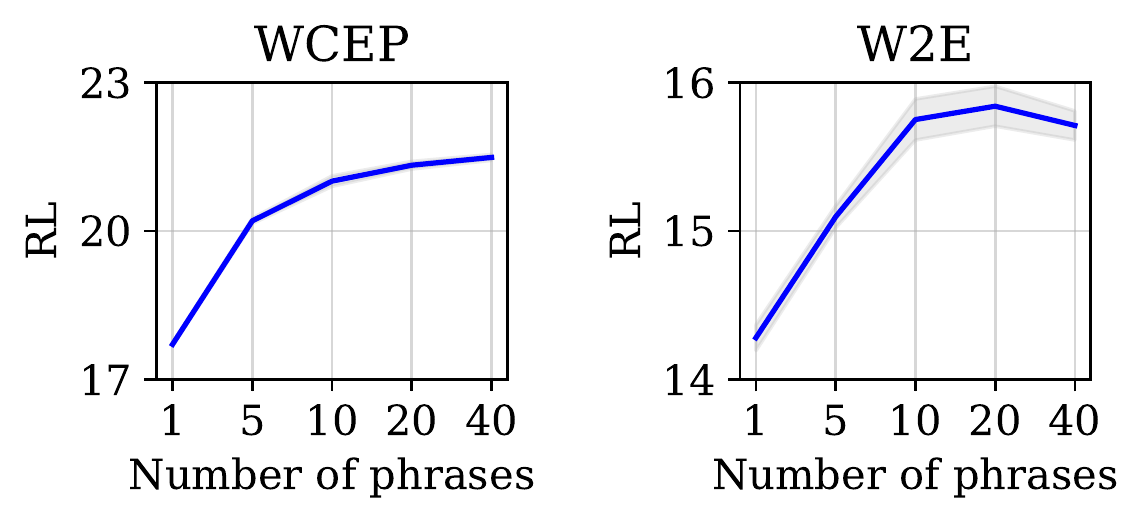}
         \vspace{-0.7cm}
        \caption{Relevance (lexical).}
        \label{fig:N_relevance}
     \end{subfigure}
     \hfill
     \begin{subfigure}[b]{0.32\textwidth}
         \centering
         \includegraphics[width=\textwidth]{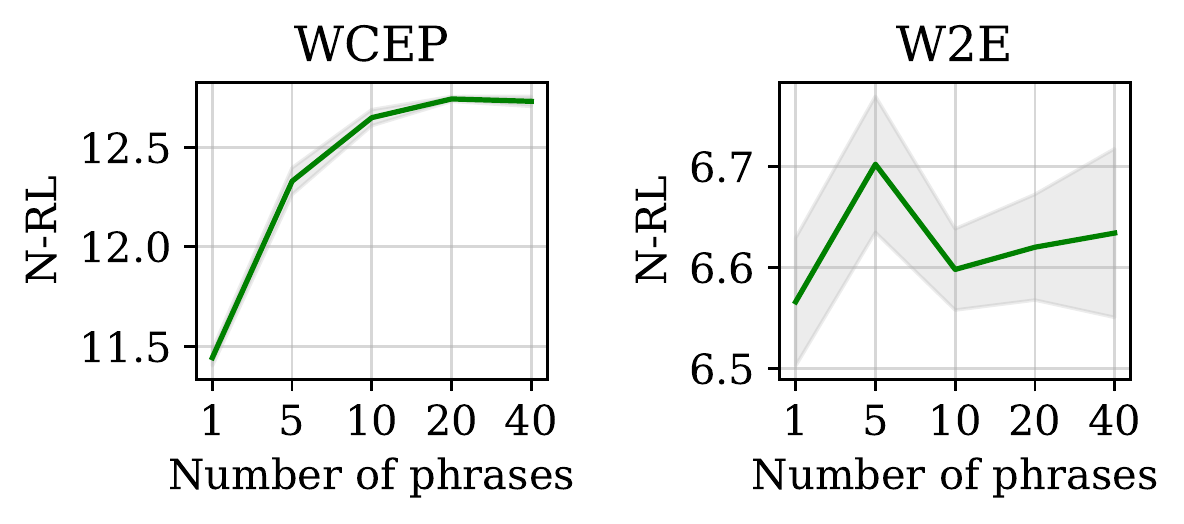}
         \vspace{-0.7cm}
        \caption{Novelty (lexical).}
        \label{fig:N_novelty}
     \end{subfigure}
     \hfill
     \begin{subfigure}[b]{0.32\textwidth}
         \centering
         \includegraphics[width=\textwidth]{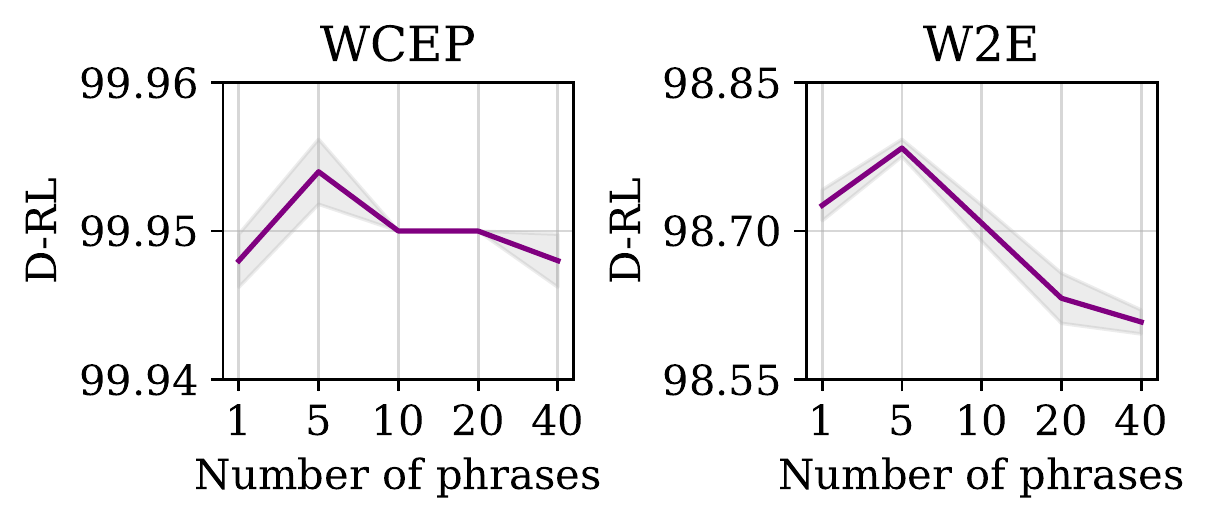}
         \vspace{-0.7cm}
        \caption{Distinctiveness (lexical).}
        \label{fig:N_distinctiveness}
     \end{subfigure}     
     \vspace{-0.3cm}
    \caption{Effects of the number of set phrases.}
    \vspace{-0.5cm}
    \label{fig:num_phrases}
\end{figure*}

\begin{figure*}[!t]
     \centering
     \begin{subfigure}[b]{0.32\textwidth}
         \centering
         \includegraphics[width=\textwidth]{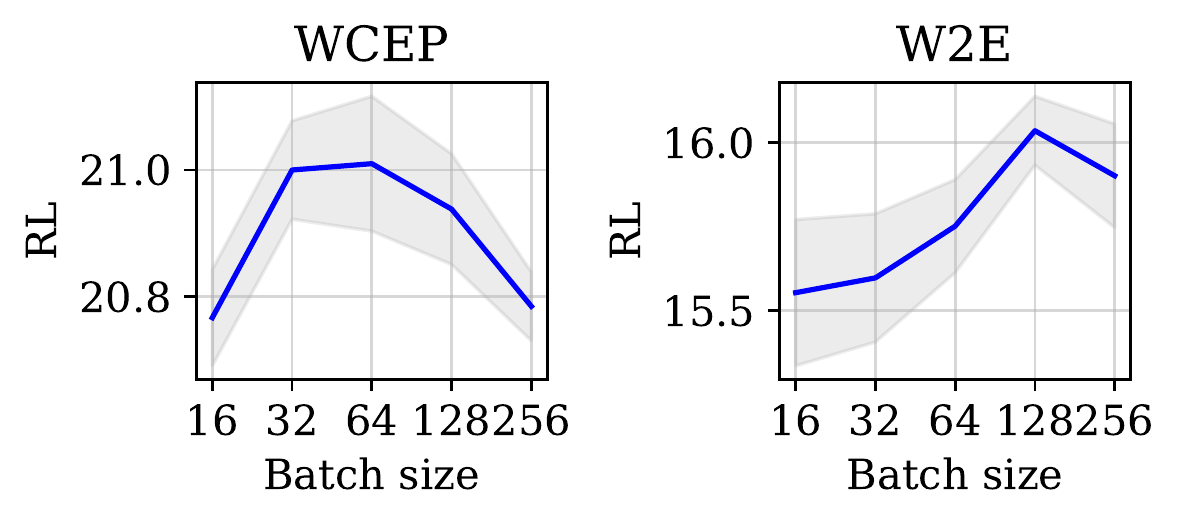}
         \vspace{-0.7cm}
        \caption{Varying batch sizes.}
        \label{fig:varying_batch}
     \end{subfigure}
     \hfill
     \begin{subfigure}[b]{0.32\textwidth}
         \centering
         \includegraphics[width=\textwidth]{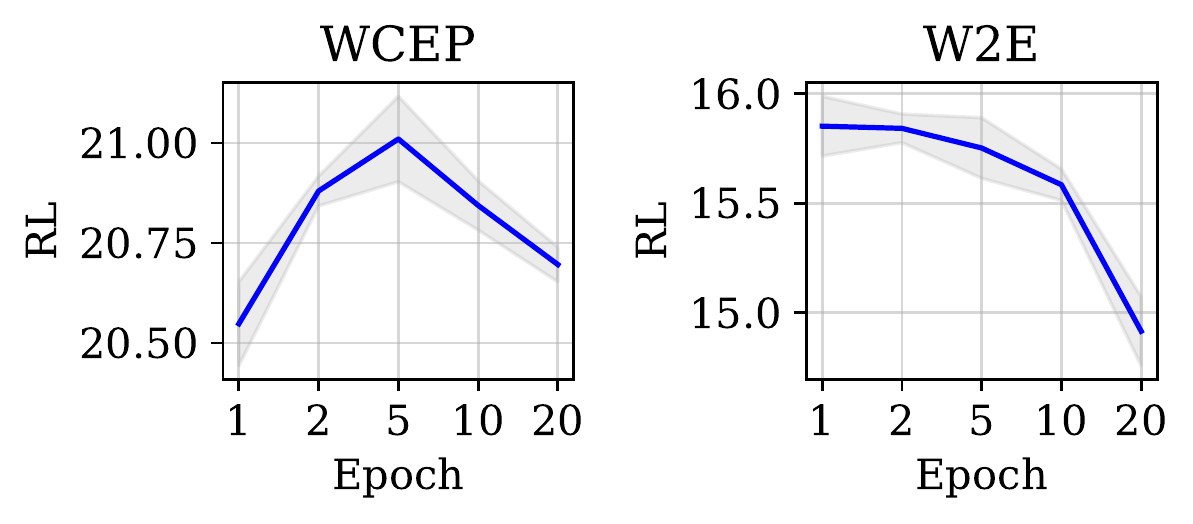}
        \vspace{-0.7cm}
        \caption{Varying epochs.}
        \label{fig:varying_epoch}
     \end{subfigure}
     \hfill
     \begin{subfigure}[b]{0.32\textwidth}
         \centering
         \includegraphics[width=\textwidth]{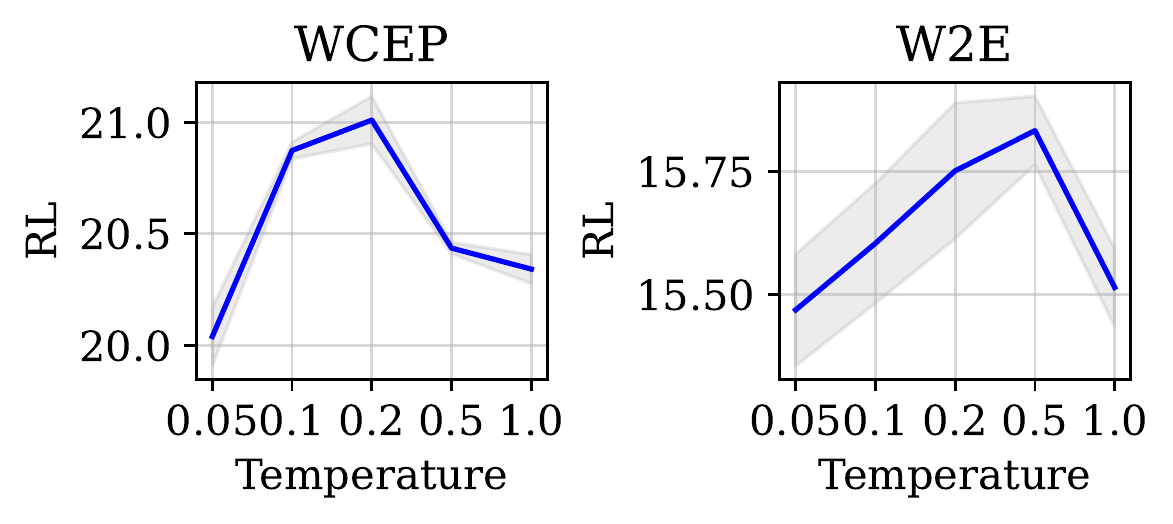}
        \vspace{-0.7cm}
        \caption{Varying temperatures.}
        \label{fig:varying_temp}
     \end{subfigure}     
     \vspace{-0.3cm}
    \caption{Sensitivity analysis on training hyperparameters.}
    \vspace{-0.5cm}
    \label{fig:hyperparam}
\end{figure*}

\begin{figure}[!t]
     \centering
     \begin{subfigure}[b]{\columnwidth}
         \centering
         \includegraphics[width=\textwidth]{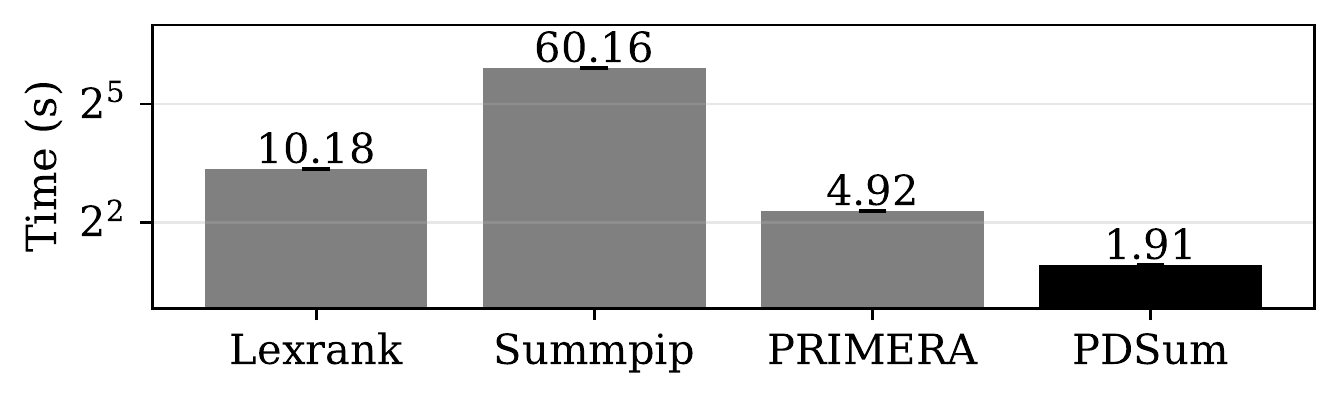}
         \vspace{-0.7cm}
        \caption{Average running time in each context.}
        \label{fig:WCEP_time}
     \end{subfigure}
     \begin{subfigure}[b]{\columnwidth}
         \centering
         \includegraphics[width=\textwidth]{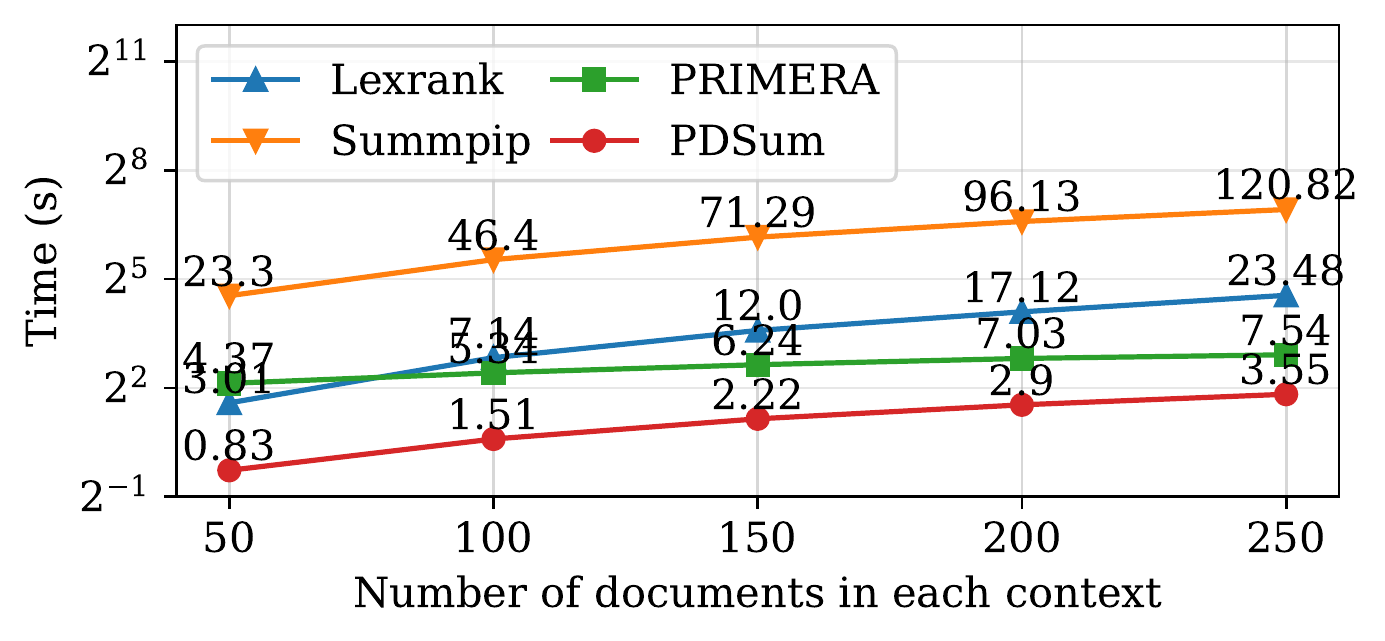}
         \vspace{-0.7cm}
        \caption{Average running time by varying the number of documents.}
        \label{fig:WCEP_scalability}
     \end{subfigure}
     \vspace{-0.7cm}
     \caption{Scalability analysis results in WCEP.}
     \label{fig:scalability_analysis}
     \vspace{-0.2cm}
\end{figure}

\subsection{Sensitivity Analysis}
\label{sec:sensitivity}
\subsubsection{\textbf{Knowledge Distillation Ratio}} Figure \ref{fig:WCEP19_RR} shows the effects of distillation ratio in WCEP (refer to Appendix \ref{apx:W2E_RR} for more discussion with the results in W2E). A novel ratio (i.e., the ratio of novel tokens in a new summary) decreases as a distillation ratio increases (i.e., \algname{} weighs more on the accumulated knowledge in previous contexts). A high novel ratio also leads to higher relevance scores to a reference summary since it could preserve the lifelong theme of a set. Interestingly, the lexical novelty score is positively correlated to the distillation ratio, while the semantic novelty score is not. This is because the novel tokens in a new summary become more conforming to the tokens in the reference summary with more preserved knowledge, but their collective semantics become less similar to the semantics of the reference summary with fewer tokens. Both of the scores for distinctiveness, however, are positively correlated with the distillation ratio which again shows preserving previous knowledge helps the distinctiveness of new summaries. Regardless of the trade-offs in various aspects of the output summaries discussed, the distillation ratio of 0.5 smoothly balanced the trade-offs and lead to the quality results in both datasets.

\subsubsection{\textbf{Number of Set Phrases}} Figure \ref{fig:num_phrases} shows the effects of the number of set phrases. We show the results of RL for brevity, while those of other metrics showed similar trends. Overall, considering more set phrases helps get more relevant and novel summaries but less distinct summaries in both datasets, which is expected as more phrases may help specify the theme of each set but also can overlap over concurrent sets. It is empirically observed that a moderate number of phrases, around five to ten, is the optimal value balancing the trade-off, which is also consistent with real-world practices (e.g., it is common to use around five keywords to describe a certain theme in news articles, scientific papers, or products).  

\subsubsection{\textbf{Training Hyperparameters}} We studied the effects of main hyperparameters used for training \algname{}: the batch size, the number of epochs, and the temperature value. Due to space limitation, in Figure \ref{fig:hyperparam}, we show the results of relevance (i.e., RL, where BS showed similar trends) which is the most important criterion in the summarization task. For both datasets, each hyperparameter has an optimal point for the highest scores, which conforms to the default value used in \algname{}. However, overall, \algname{} consistently achieved good performances as even its lowest scores in extreme settings were higher than the best score of existing methods (e.g., the best RL score by PRIMERA was 18.45 in WCEP and 14.24 in W2E as reported in Table \ref{tbl:overall_performance}). This again demonstrates the merits and robustness of \algname{} over various settings in EMDS.

\subsection{Scalability Analysis}
\label{sec:scalability} Besides the summarization quality, the algorithms for EMDS should be efficient and scalable to deal with dynamic streaming environments. We measured the average running time of compared algorithms in summarizing sets in each context. As shown in Figure \ref{fig:scalability_analysis} for WCEP (the results for W2E showed similar trends), \algname{} was the fastest among the compared algorithms by taking only a few seconds to summarize hundreds of documents. This is attributed to lightweight prototype-based sets processing while the existing algorithms not only process each set independently but also use expensive graph-based processing (Lexrank and Summpip) or a large language model (PRIMERA). Furthermore, when we varied the input rate of an evolving multi-document sets stream by controlling the number of documents in each context, \algname{} consistently achieved the lowest running time with a comparable increase rate. The trend of scalability also conforms to the time complexity of \algname{} which is linear to the number of documents.
\smallskip
\vspace{0.2cm}
\section{Discussion and Conclusion}
Before concluding, we discuss two interesting directions to facilitate future work for EMDS. First, an abstractive summarization approach can be alternatively considered. For instance, in \algname{}, the summary identifier can incorporate a decoding module that considers a set prototype as a key signal to decode the output summary from documents. However, to prevent the hallucination problem, the factual consistency over different contexts in the same set needs to be specifically considered in accumulating previous knowledge and generating summaries. Second, while an unsupervised approach is practical in EMDS, some information can be provided as delayed feedback such as reference summaries by annotators or user ratings on new summaries. Then, they can be used to refine the embedding and summarizing processes as previous auxiliary knowledge (e.g., to additionally regularize set prototypes or sentence score functions in \algname{}).

In conclusion, we introduced a new summarization task EMDS for continuously summarizing evolving multi-document sets stream. We proposed a novel unsupervised method \algname{} for EMDS, that builds \emph{lightweight prototypes of multi-document sets} used for embedding and summarizing. The relevance, novelty, and distinctiveness of summaries are achieved by continuously updating set prototypes over contexts through a contrastive learning objective, while being regularized by accumulated knowledge distillation. We demonstrated the superiority of \algname{} over existing state-of-the-art unsupervised summarization algorithms in benchmark datasets. We believe this work opens a promising direction for summarization. 


\clearpage
\begin{acks}
The first author was supported by Basic Science Research Program through the National Research Foundation of Korea (NRF) funded by the Ministry of Education (2021R1A6A3A14043765). The second author was supported by the Science and Technology Development Fund, Macau SAR (Grant No. 060/2017/AFJ and 070/2022/AMJ). The research was supported in part by US DARPA KAIROS Program No. FA8750-19-2-1004 and INCAS Program No. HR001121C0165, National Science Foundation IIS-19-56151, IIS-17-41317, and IIS 17-04532, and the Molecule Maker Lab Institute: An AI Research Institutes program supported by NSF under Award No. 2019897, and the Institute for Geospatial Understanding through an Integrative Discovery Environment (I-GUIDE) by NSF under Award No. 2118329. The views and conclusions contained in this paper are those of the authors and should not be interpreted as representing any funding agencies.
\end{acks}


\bibliographystyle{ACM-Reference-Format}

\balance
\bibliography{reference}

\clearpage
\appendix
\section{Supplemental Material}
\subsection{Details of Datasets }
\label{apx:dataset}
\renewcommand{\arraystretch}{1}
\begin{table}[!h]
\small
\vspace{-0.2cm}
\caption{Statistics of the datasets.}
\vspace{-0.4cm}
\label{tbl:datasets}
\begin{tabular}{cccccc}
\toprule
Dataset    & Period                     & \#Docs & \#Sets & \#RefSummaries\\ \hline
WCEP\,\cite{WCEP}     & Jan$\sim$Dec, 2019 & 29,931                & 519 & 519      \\
W2E\,\cite{W2E}     & Jan$\sim$Dec, 2016 & 14,475
& 47 & 248      \\
\bottomrule
\end{tabular}
\vspace{-0.3cm}
\end{table}
The statistics of two datasets are summarized in Table \ref{tbl:datasets}.
\begin{itemize}[leftmargin=10pt, noitemsep]
    \item WCEP\,\cite{WCEP} is a large-scale benchmark data set with reference summaries. We chose news stories with at least 50 news articles over their whole lifespan and used the news articles of the stories published in 2019. We simulated an evolving multi-document set stream by feeding articles ordered by their publication date, thus each day becomes a context for summarization. As each story has a single reference summary, it is used for evaluation in each context of the story.
    \item W2E\,\cite{W2E} is another large-scale benchmark data set. As it provides multiple reference summaries spanned over the lifespan of news stories, for each story, we regarded the temporal gaps between consecutive reference summaries as the contexts (i.e., to compare with a reference summary at $t$, a new summary is generated from the articles published from $t\!-\!1$ to $t$). We used news stories including at least 10 articles in each context. Similar to WCEP, we simulated an evolving multi-document set stream by feeding articles ordered by their publication date. We conducted summarization in each unique context in the stream, where the overlapped temporal period among the contexts naturally leads to concurrent articles and sets in the context.
\end{itemize}

\subsection{Implementation of Compared Algorithms}
\label{apx:implementation}
\begin{itemize}[leftmargin=10pt, noitemsep]
    \item \emph{Variants of centroid-based model}\,\cite{centroid_revisit,centroid_word, MEAD}: Instead of using symbolic\,\cite{tfidf} or shallow\,\cite{w2v} approach for embedding sentences and documents, we used a deep contextualized embedding with a pretrained language model (sentence-BERT\,\cite{sentencebert}) which shows the state-of-the-art performances in semantic tasks.
    \begin{itemize}[leftmargin=10pt, noitemsep]
        \item \emph{SentCent} and \emph{DocCent}: For each context, the center of each set is calculated by the mean of sentence (or document) representations and the sentences closest to the center are returned.
        \item \emph{IncSentCent} and \emph{IncDocCent}: Similar to the above, but the centers are incrementally updated from the previous contexts.
    \end{itemize}
    \item \emph{Lexrank}\,\cite{lexrank}: A widely used unsupervised graph-based summarization algorithm that uses centrality scores based on PageRank\,\cite{pagerank} for ranking sentences. We used the default settings following the original work.
    \item \emph{Summpip}\,\cite{summpip}: A state-of-the-art unsupervised extractive summarization algorithm with graph clustering and compression. We used sentence-BERT\,\cite{sentencebert} for embedding sentences and the default settings following the original work.
    \item \emph{PRIMERA}\,\cite{primera}: A state-of-the-art abstractive summarization algorithm with self-supervision by the Gap Sentence Generation and the Entity Pyramid. We used a pretrained model provided by the authors trained with a large-scale news data set (Newshead\,\cite{newshead}) without reference summaries (i.e., unsupervised).
    \item \textbf{\emph{\algname{}}} (proposed): We used TFIDF\,\cite{tfidf} for a phrase ranker and sentence-BERT\,\cite{sentencebert} for initializing sentence representations. We set the number of heads for MHS to 2, the hidden dimensionalities to 1024, the learning rate to 1e-5, the batch size to 64, the number of epochs to 5, the temperature value to 0.2, and the distillation ratio to 0.5. We reported the mean and standard error of the scores measured from ten independent experiments (the other algorithms, including a pretrained PRIMERA, are deterministic).
\end{itemize}
For sentence-BERT\,\cite{sentencebert}, we used a popular pretrained model \emph{all-roberta-large-v1}. For existing algorithms, the documents in each set in each context are used as the input for summarization. For all algorithms, we set the size of an output summary to one sentence (for extractive approaches) or 40 tokens (for abstractive approaches), following the average size of reference summaries in the datasets used. All experiments are conducted with a 2.5GHz 32-Core CPU with 1TB RAM and an RTX A6000 GPU with 48GB RAM.

 \begin{figure*}[!t]
     \centering
     \includegraphics[width=\textwidth]{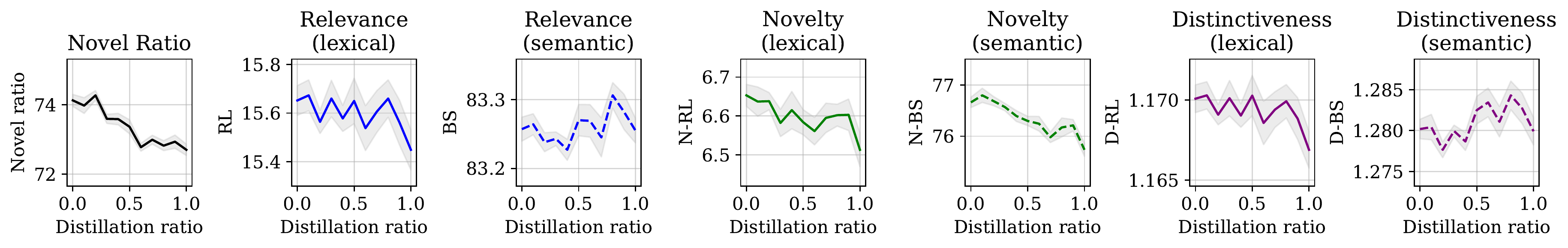}
     \vspace{-0.7cm}
    \caption{Effects of knowledge distillation ratio in W2E.}
    \vspace{-0.3cm}
    \label{fig:W2E_RR}
 \end{figure*}

\begin{figure*}[!t]
     \centering
     \begin{subfigure}[b]{\textwidth}
        \centering
        \includegraphics[width=\textwidth]{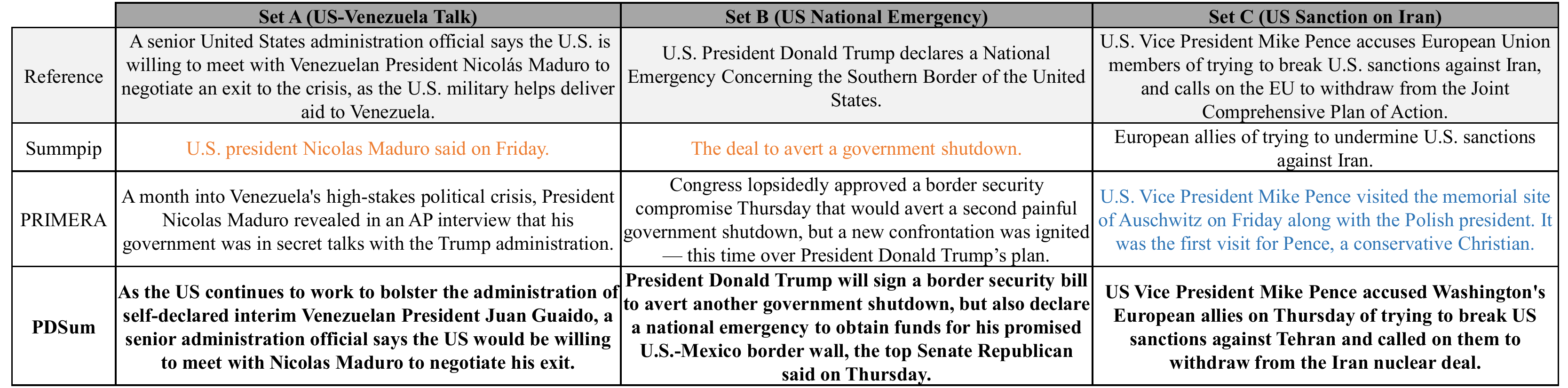}
        \vspace{-0.5cm}
        \caption{Summaries of \emph{three different sets} of articles about \texttt{the United States} published on the same date in WCEP.}
        \vspace{0.1cm}
        \label{fig:casestudy_sets}
     \end{subfigure}
     \hfill
     \begin{subfigure}[b]{\textwidth}
        \centering
        \includegraphics[width=\textwidth]{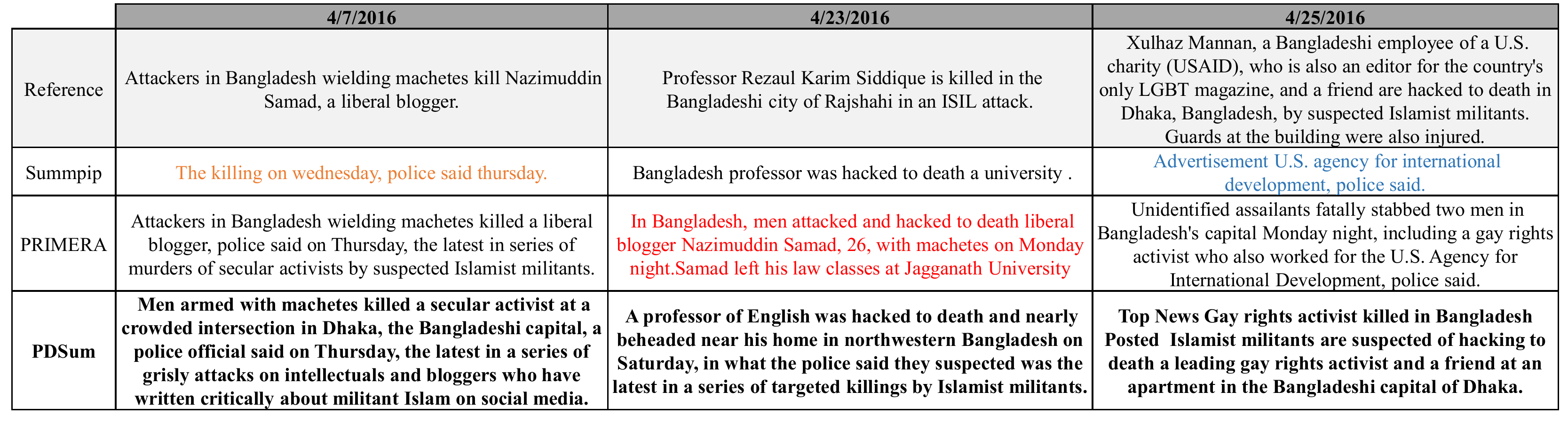}
        \vspace{-0.5cm}
        \caption{Summaries \emph{over three different days} on a set about \texttt{Attacks on secularists in Bangladesh} in W2E.}
        \label{fig:casestudy_days}
     \end{subfigure}
     \vspace{-0.7cm}
     \caption{Example summaries of compared algorithms, where {\color{orange} abstract}, {\color{blue} irrelevant}, or {\color{red} redundant} ones are highlighted in colors.}
     \label{fig:casestudy}
     \vspace{-0.2cm}
\end{figure*}

\subsection{Details of Evaluation Criteria}
\label{apx:evaluation}
Given an output summary $\mathbb{S}$ and a reference summary $\mathbb{G}$, let $\mathcal{F}(\mathbb{S},\mathbb{G})$ be an evaluation metric (e.g., ROUGE\,\cite{rouge} or BERTScore\,\cite{bertscore} in our evaluation). Then, we derive the three measures with $\mathcal{F}(\cdot,\cdot)$ for evaluating EMDS by different criteria as follows. 

Given an output summary $\mathbb{S}_i^T$ for a set $i$ in a context $T$ and the corresponding reference summary $\mathbb{G}_i^T$,
\begin{itemize}[leftmargin=10pt, noitemsep]
    \item \textbf{\emph{Relevance}} measures the default score by $\mathcal{F}$:
    \begin{equation}
    \small
    \text{Score}_{relevance} = \mathcal{F}(\mathbb{S}_i^T, \mathbb{G}_i^T)
    \end{equation}
    \item \textbf{\emph{Novelty}} measures how relevant the novel part of a new summary from the previous summary is to the reference summary:
    \begin{equation}
    \small
    \text{Score}_{novelty} = \mathcal{F}(\mathbb{S}_i^T \setminus \mathbb{S}_i^{T-1}, \mathbb{G}_i^T),
    \end{equation}    
    where $\mathbb{S} \setminus \mathbb{S}^{\prime}$ denotes the token-level differences.
    \item \textbf{\emph{Distinctiveness}} measures how distinctive a new summary is from other reference summaries. Specifically, the distinctiveness of a summary from other reference summaries is calculated by $1-\mathcal{F}(\mathbb{S},\mathbb{G})$ and normalized by that from its reference summary (i.e., the higher the distinctiveness score, the less similar an output summary is to other reference summaries):
    \begin{equation}
    \small
    \text{Score}_{distinctiveness} = \frac{\frac{1}{|\mathcal{G}^T|-1}\sum_{\mathbb{G}_j^T \in \mathcal{G}^T, j \neq i}(1-\mathcal{F}(\mathbb{S}_i^T, \mathbb{G}_j^T))}{1-\mathcal{F}(\mathbb{S}_i^T, \mathbb{G}_i^T)},
    \end{equation}
    where $\mathcal{G}^T$ is a collection of reference summaries in all sets in $T$.
\end{itemize}

\subsection{Knowledge Distillation in W2E}
\label{apx:W2E_RR}
Following the effects of knowledge distillation ratio in WCEP discussed in Section \ref{sec:sensitivity}, Figure \ref{fig:W2E_RR} shows the results in W2E. Unlike the trends observed in WCEP, as the distillation ratio increases, the lexical score in each criterion is consistent or marginally decreases while the semantic score is also consistent or marginally increases (except for novelty). This is because in W2E output summaries are evaluated with individual reference summaries in each context, which may have different thematic keywords (i.e., thus not necessarily ''lexically'' relevant) but should have some degree of similar semantics (i.e., as they are about the same set). In the case of the semantic novelty score, however, considering the previous knowledge more should make the overall semantics of a new summary less similar to that of a reference summary, as conforming to the observation in WCEP. Overall, the default distillation ratio of 0.5 balances the trade-off well, showing competitive scores in most cases in W2E, as well as in WCEP.

\subsection{Qualitative Analysis}
\label{apx:case_study}

\begin{table}[!t]
\vspace{-0.2cm}
\caption{Human evaluation results.}
\label{tbl:human_eval}
\vspace{-0.4cm}
\begin{tabular}{ccccccc}
\toprule
    & \multicolumn{2}{c}{Relevance}  & \multicolumn{2}{c}{Novelty} & \multicolumn{2}{c}{Distinctiveness}  \\ 
 & WCEP & W2E & WCEP & W2E & WCEP & W2E \\ \hline
Summpip     & 1.56 & 1.08 & 1.80  & 1.80 & 1.80 & 1.73     \\
PRIMERA     & 2.44 & 2.16 & 1.88 & 2.76 & 3.40 & 3.07      \\
\algname{}     & \textbf{4.08} & \textbf{3.68} & \textbf{3.40} & \textbf{3.64} & \textbf{3.60} & \textbf{3.93}      \\
\bottomrule
\end{tabular}
\vspace{-0.1cm}
\end{table}

We conducted a qualitative analysis of \algname{} in comparison with the two existing algorithms based on extractive (Summpip) and abstractive (PRIMERA) approaches, respectively. For human evaluation, we recruited five graduates to evaluate the compared algorithms' summaries for 25 news stories of different categories (e.g., Politics, Sports, etc.) in the two datasets. The raw articles and reference summaries were provided together. Following the widely used protocol\,\cite{summeval}, we asked them to rate summaries on a Likert scale from 1 to 5 (i.e., from very poor to excellent) for relevance, novelty, and distinctiveness. As shown in Table \ref{tbl:human_eval}, \algname{} achieved higher scores than the existing algorithms by a large margin.

We further analyzed the quality of summaries by each algorithm with two case studies. Figure \ref{fig:casestudy} shows the case studies on example summaries generated by \algname{} and the two existing algorithms. Figure \ref{fig:casestudy_sets} shows the case study in WCEP, where the summaries of the three \emph{different sets in the same context} with similar themes about \texttt{the United States} are provided. The existing algorithms were not effective in returning relevant and distinctive summaries for different sets, by including too abstract or irrelevant summaries. On the other hand, the summaries returned by \algname{} were distinctive from each other as well as relevant to the reference summaries. Figure \ref{fig:casestudy_days} shows the case study in W2E where summaries of the \emph{same set in different contexts} are provided. Similarly, on each day, \algname{} could get the consistently relevant summaries to the set about \texttt{Attacks on secularists in Bangladesh}, while being specific to each accident that happened each day. The two existing algorithms, however, returned too abstract or irrelevant summaries (Summpip), without capturing the relevance and distinctiveness effectively, or a redundant summary (PRIMERA) without capturing the novelty effectively.

\end{document}